\documentclass[useAMS,usenatbib]{mn2e}

\usepackage{savesym}
\usepackage{graphicx}
\usepackage{fixltx2e} % for superscript, subscript text: \textsuperscript{} & \textsubscript{}
\usepackage{dcolumn}
\usepackage[version=3]{mhchem} % for chemical formula: e.g. \ce{H2O}, \ce{C^{18}O}
\usepackage{natbib}
\usepackage{color}
\usepackage{latexsym}

\newcommand{\kms}{km\,s$^{-1}$}

\newcommand{\solarM}{M$_\odot$}

\title[Scaled up LMSF in massive SF cores]{Scaled up low-mass star formation in massive star-forming cores in the G333 giant molecular cloud}

\author[B. Wiles et al.]{B.~Wiles$^{1}$, N.~Lo$^{2}$, M. P.~Redman$^{1}$\thanks{matt.redman@nuigalway.ie}, M. R.~Cunningham$^{3}$, P. A.~Jones$^{2,3}$, M. G.~Burton$^{3}$ \and L.~Bronfman$^{2}$\\
  $^{1}$Centre for Astronomy, School of Physics, National University of Ireland Galway, University Road, Galway, Ireland\\
  $^{2}$Departamento de Astronom\'ia, Universidad de Chile, Camino El Observatorio 1515, Las Condes, Santiago, Casilla 36-D, Chile\\
  $^{3}$School of Physics, University of New South Wales, Sydney 2052, Australia\\
}

\begin{document}

\date{DRAFT}

\pagerange{\pageref{firstpage}--\pageref{lastpage}} \pubyear{}

\maketitle

\label{firstpage}

\begin{abstract}
Three bright molecular line sources in G333 have recently been shown to exhibit signatures of infall. We describe a molecular line radiative transfer modelling process which is required to extract the infall signature from Mopra and Nanten2 data. The observed line profiles differ greatly between individual sources but are reproduced well by variations upon a common unified model where the outflow viewing angle is the most significant difference between the sources. The models and data together suggest that the observed properties of the high-mass star-forming regions such as infall, turbulence, and mass are consistent with scaled-up versions of the low-mass case with turbulent velocities that are supersonic and an order of magnitude larger than those found in low-mass star-forming regions. Using detailed radiative transfer modeling, we show that the G333 cores are essentially undergoing a scaled-up version of low mass star formation. This is an extension of earlier work in that the degree of infall and the chemical abundances are constrained by the RT modeling in a way that is not practical with a standard analysis of observational data. We also find high velocity infall and high infall mass rates, possibly suggesting accelerated collapse due to external pressure. Molecular depletion due to freeze-out onto dust grains in central regions of the cores is suggested by low molecular abundances of several species. Strong evidence for a local enhancement of \ce{^{13}C}-bearing species towards the outflow cloud cores is discussed, consistent with the presence of shocks caused by the supersonic motions within them.
\end{abstract}


\section{Introduction}

Massive star formation must eventually differ substantially from low mass star formation because the outcomes are so different. Massive stars form in groups inside dense clusters of lower mass stars and whereas in low mass star formation, more isolated single stars or binary stars are formed. Both modes of star formation are initiated in cold molecular clouds so a key question is at what point of evolution do the two types of star formation diverge.

For several decades, high quality data have been available for nearby low mass star formation. Spatially resolved molecular line observations in many species and transitions can be readily obtained. By combining this data with radiative transfer codes, it has proved possible to extract quantitative dynamical information from these high quality line profiles. In contrast, massive star formation is intrinsically rarer and of shorter duration than low mass star formation which makes observational targets statistically much more distant. However, comparable quality data to that of low mass star formation is now becoming available. Major observational spectral line surveys in the millimetre regime are now being carried out of massive star forming regions across the galaxy, for example in NH$_3$ \citep[HOPS,][]{walsh11}, CO \citep{burton13}, CS \citep{jordan15}, and multiple lines MALT90 \citep{jackson13}.  

G333 is one of the most massive GMCs in the 4th quadrant of the Galaxy and has been characterised by \citet{garcia14} as being of radius 67 pc, distance 3.6 Kpc and mass $2\times 10^6~{\rm M_\odot}$. In a companion paper to this work, Lo et al 2015 \nocite{LoObs} present data from an observational survey of G333 on the Nanten and Mopra telescopes. This work is part of a series of observational studies of this region \citep{Lo2011,Lo2009,wong08,bains07,Lo2007}. The data revealed that three bright molecular line sources exhibit signature signs of infall. These are designated G333.6--0.2, G333.1--0.4 and G332.8--0.5. Within a Mopra beam size of the outflow sources are radio-detected H{\sc ii} regions, 1.2-mm dust emission clumps \citep{Mookerjea2004}, \ce{CH3OH} and \ce{H2O} masers and {\it IRAS} sources, features which are all consistent with the presence of high mass star formation. The mass and temperature of each source were confined with a detailed analysis of the observed luminosities. In particular, the masses were constrained by using the \ce{^{13}CO}(1--0) line flux as the CO(1--0) lines were found to be optically thick. Therein results from a first attempt at radiative transfer modelling of the sources were presented. Here we expand on the findings from the observations and earlier modelling by presenting radiative transfer modelling of the three sources in agreement with the observational results to quantify limits for the turbulent velocity, infall velocities, and chemical abundances.

%introduce clouds, messy fields, HCO+ and CO data

As the three high mass star forming regions (HMSFR) are located within a massive giant molecular cloud (GMC) complex, the analysis was hampered by the multiplicity of sources present and the confusion of their association with features both spatially and kinematically. The outflows are thus less well-defined than in low mass star forming regions (LMSFR) and the clean, detailed analysis that is possible there is not possible here.  In particular, G333.6--0.2 presents a complex field of features in CO(1--0) and presents very strong absorption in \ce{HCO+}(1--0) and so also presents a particular challenge for modelling.
Despite these complications, many physical parameters can be constrained from the observations. These data and their analysis are presented in full in \citet{LoObs}. Here we reproduce the tables summarizing the major results in Table~\ref{tab:derivedproperties}. We will constrain ourselves to models that are in agreement with the core and outflow properties determined by \citet{LoObs}.

\subsection{Observations of G333}
The observational evidence presented by \citet{LoObs} for outflows from the three sources is robust. Outflows are ultimately powered by infall but the dynamical signature of infall is much harder to detect for several reasons. Firstly, the gas undergoing the most rapid infall is located close to the central protostellar source-disk system which is typically unresolved by single dish observations. Gas in the outer regions of the cloud may have small infall motions but these can easily be masked by effects such as turbulence or bulk oscillations of the cloud. Secondly, infall takes place over a spherical volume in contrast to the highly directional outflows so that gas moving in a range of directions is typically present within a beam. Thirdly, effects such as rotation \citep{Redman04}, turbulence and bulk oscillations \citep{Alves01,Redman2006} can complicate the identification of an infall signature. To demonstrate that infall is also present therefore requires careful consideration of the radiative transfer effects taking place in potential sources.

\begin{figure*}
  \includegraphics[width=1.0\textwidth]{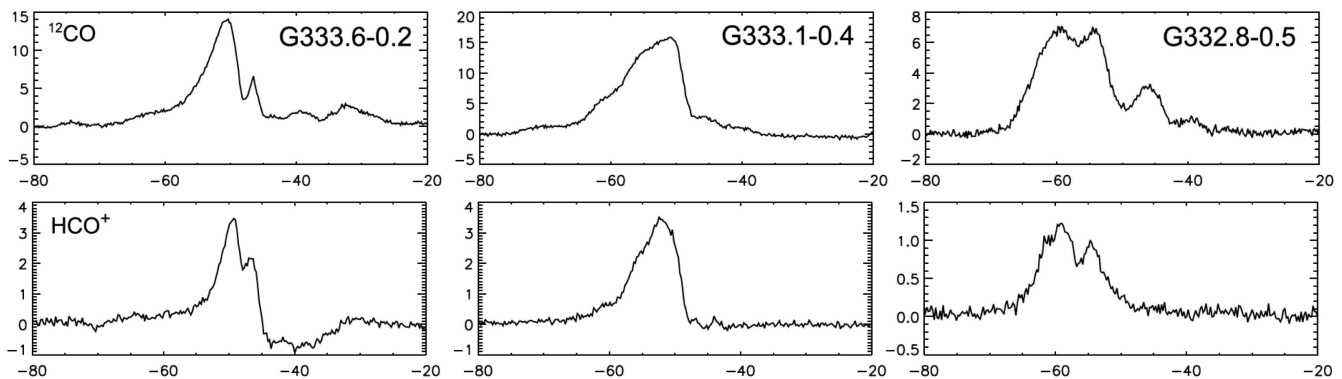}
  \caption{Example velocity profiles of the molecular lines for each of the outflow sources. The spectra were taken from central positions of the observed cores. Left: G333.6--0.2; centre: G333.1--0.4; right: G332.8--0.5.  The top row is CO spectra and the bottom row is \ce{HCO+} spectra. Both lines are from the $J$ = 1 -- 0 transition. The y-axis is the intensity, $T_{\rm A}^*$, in units of K and the x-axis is the $v_{\rm lsr}$ velocity in \kms.  For details of the full spectral dataset, including lines of H$^{13}$CO$^+$, N$_2$H$^+$, and CO isotopologues see \citet{LoObs}.}
  \label{fig:profiles}
\end{figure*}
\begin{table*}
  \centering
  \caption{Molecular lines observed. Columns are as follows: (1) molecule, (2) transition; (3) rest frequency; (4) months for observations; (5) velocity resolution; (6) typical $1 \sigma$ rms off-line noise level per velocity channel in terms of measured $T_{\rm A}^*$; (7) reference. Reproduced from \citet{LoObs}}
  \label{tab:obsdetails}
  \begin{tabular}{@{}lcccccc}
    \hline
    Molecule & Transition & Rest frequency & Months Observed & $\Delta v$ & $1 \sigma$ & Reference \\
     &	       & GHz	    &	   &	\kms	  &	K	   &	\\
    (1)	 &	(2)    & (3)	    &	(4)	   &	(5)	  &	 (6)   &	(7) \\
    \hline
    \ce{^{12}CO} &	$J=1-0$ & 115.27 &	2006 Aug	   &	0.09	  &	0.2	 & \citet{LoObs} \\
    \ce{^{13}CO} &	$J=1-0$ & 110.20  &	2004 Jun - Oct &	0.17	  &	0.1	 & \citet{bains06} \\
    \ce{C^{18}O} &	$J=1-0$ & 109.78 &	2005 Jul - Sep &	0.17	  &	0.1	 & \citet{wong08} \\
    \ce{CS}		 &	$J=2-1$ &  97.981 &	2006 Sep - Oct &	0.10	  &	0.1	 & \citet{Lo2009} \\
    \ce{HCO+}	 &	$J=1-0$ &  89.190 &	2006 Jul - Sep &	0.11	  &	0.1	 & \citet{Lo2009} \\
    \ce{H^{13}CO+} &$J=1-0$ &  86.754 &	2006 Jul - Sep &	0.12	  &	0.2	 & \citet{Lo2009} \\
    \hline
  \end{tabular}
\end{table*}

\begin{table}
  \centering
  \caption{Derived core and outflow parameters, reported by Lo et al 2015. The columns are as follows: (1) source; (2) $M_{\rm core}$ Core mass derived from dust continuum emission;  (3) the sum of the blue and red outflow masses; (4) $l$ is the outflow length scale; (5) $t_{\rm outflow}$ outflow time for this scale size; (6) $\dot{M}_{\rm loss}$ the mass loss rate derived for each outflow (red + blue components)}
  \label{tab:derivedproperties}
  \begin{tabular}{@{}ccccccc}
    \hline
    Source& $M_{\rm core}$ &$M_{\rm outflow}$ & $l$ & $t_{\rm outflow}$  & $\dot{M}_{\rm loss}$  \\
     & \solarM &  \solarM & pc & $10^{3}$\,yrs &\solarM/yr \\
    (1) & (2) & (3) & (4) & (5) & (6) \\
    \hline
    G333.6--0.2 & 3300 & 68 & 0.17 & 10 & $6.4\times 10^{-3}$  \\
    G333.1--0.4 & 1800 & 52 & 0.50 & 40 & $1.5\times 10^{-3}$  \\
    G332.8--0.5 & 1100 & $23$ & 0.67 & 80 & $0.3\times 10^{-3}$ \\
    \hline
  \end{tabular}
\end{table}

in \citet{LoObs}, data analysis was performed on lines from the $J$ = 1--0 transition in CO, \ce{^{13}CO}, \ce{C^{18}O}, ${\rm HCO^+}$, and ${\rm H^{13}CO^+}$ and from the $J$ = 2--1 transition in CS, observed in all three cores. For CO isotopologues ($\sim115$ GHz range), beam size is $\sim30$ arcsec, for the rest ($\sim90$ GHz), beam size is 38 arcsec. The ratio of the three CO isotopologues was used to correct for optical depth at each velocity channel and thus to determine the column density for the line core as well as the outflow lobes. From this we are able to determine their masses, as well as estimate mass loss rates, outflow mechanical energies and luminosities. Typical outflow masses are 10 to 40 \solarM in each lobe, compared to core masses of order $10^3$ \solarM. Outflow size scales are a few tenths of a parsec, timescales are several $\times 10^4$ years and mass loss rates a few $\times 10^{-3}$ \solarM/yr. The source SED were used to calculate their luminosities, and then by fitting to a 2-component grey-body model, the dust mass, dust temperature and source size for the extended component were also determined.

In this paper, we apply a 3D radiative transfer analysis making use of the code {\it MOLLIE}, which is able to consider the competing contributions of the outflow, infall and ambient gas, which may also have different densities, temperatures and chemical compositions, in order to provide an estimate of the source parameters, and in particular to yield mass infall rates and infall speeds from the data set. Estimates could not simply be made from the magnitude of the line splitting on the basis of the observations alone but must consider the medium through which the radiation passes with a full radiative transfer analysis. We model the source geometry as well as its physical characteristics in order to yield line profiles in agreement with the observations.

The observations have provided us with constraints for core mass and temperature. For further details of the observations, see Table~\ref{tab:obsdetails} and references therein.  For full details of the observational analyses and results, see \citet{LoObs}. We will thus find a fit for the velocity structure, turbulence, and chemical abundance by restricting ourselves to models that reproduce the observed masses and temperatures. We rely further on the observed molecular lines as detailed in the following sections.

\section{MOLECULAR LINE RADIATIVE TRANSFER MODELLING OF INFALL} \label{sec:modelling}

In low mass star formation (LMSF), the signature of gravitational infall is a double-peaked line profile in which the blue peak is stronger than the red peak \citep[e.g.][]{Myers1996,Choi95,zhou.et.al93}. Such a profile will only emerge if the temperature, density and velocity increase radially inward in an optically thick line transition. The strongest emission from both the blue- and red-shifted infalling gas is located close to the spatial centre but the red-shifted emission is absorbed along the line-of-sight by cooler material further from the centre. The blue peak is less affected since it is Doppler-shifted away from the velocity of the bulk of the absorbing gas. Such a profile is best observed when the outflow is close to the plane of the sky \citep[e.g. B335]{hodapp98} and therefore does not interfere with the infall signature.

In high mass star formation (HMSF), it now appears that the same effects are taking place as in LMSF with the same overall line profile shape of a blue wing stronger than the red wing; see for example \citet{Wu2005} and \citet{Carolan2008}. However, in HMSF the lines are much broader overall which is caused by greater infall and turbulent velocities than in low mass star formation. As with all aspects of massive star formation, individual sources tend to suffer from confusion due to the effects of distance and multiplicity. In addition, there are several other effects that cause deviations from a clear blue asymmetric line profile. Multiplicity of sources can lead to more than one density peak at the source of the outflow (if multiple outflows are present, the most powerful strongly dominates because the outflow power scales with the mass of the driving source). Rotation or bulk turbulence can lead to infall-mimicking line signatures \citep{Redman2006}. For the massive sources investigated here, the radiation field of the ambient environment is an additional effect that needs to be considered. Embedded ultracompact H{\sc ii} regions are prodigious sources of IR- and mm-wavelength continuum radiation, which can strongly affect the observed lines from less-evolved nearby clouds along the same line-of-sight. In G333, 14 FIR sources contribute to a total luminosity of $3\times 10^6~{\rm L_\odot}$ \citep{garcia14} and there is strong evidence of the effects of this in two of the sources studied here (G333.6--0.2 and G333.1--0.4).

\subsection{Modelling process} \label{sec:mollie}
It is possible to take into account the effects described above to generate relatively robust self-consistent models of individual massive star forming cores and extract the infall signatures. We break down the analysis and modeling process here into a series of steps in an approach that has been used in previous work \citep{Rawlings2013,Lo2011,Carolan2009,Carolan2008}. To aid in the reproducibility of our work, and to serve as a suggested approach for others, we now describe the following steps:\\
i) Define source geometry.\\
ii) Pre-constrain density and temperatures from continuum observations.\\
iii) Identify tightly constrained parameters from line observations.\\
iv) Fit the few remaining parameters to match line profiles using a radiative transfer code.\\
v) Determine source physical properties from best fit models.

%We address each step in turn in the following sections.

\subsubsection{Source geometry}
The cloud size is readily determined by spatially resolved maps of the individual sources described in \citet{LoObs} and is typically $\sim 1~{\rm pc}$ and to first order, the clouds can be described as spherical. As described earlier, a star forming core undergoing gravitational collapse will always lead promptly to a disk and jet-powered bipolar outflow. The disk and any nearby rotating material will not be resolved by the single dish observations described here, and can be ignored (they will be resolved in the future by ALMA). A range of possible plausible outflow morphologies may be envisaged, ranging from a narrow pencil-beam to a wide angled cone. However, observationally from both low mass analogues at the late class 0 stage \citep[e.g.]{jorgensen.et.al07} and from GLIMPSE imagery presented in \cite{LoObs} it appears that the outflows at this stage have a characteristic hourglass type morphology with a broad opening angle and then a recollimation to a more tightly collimated outflow. Following \citet{Rawlings2004,Rollins14,Carolan2009,Carolan2008} it is convenient to describe this morphology with the following simple function:
\begin{equation}
z= \tanh(\psi r)
\label{eq:layer}
\end{equation}
where $z$ and $r$ are the coordinates along and perpendicular to the outflow axis respectively. Thus, $\psi$  defines the overall shape of the outflow. A value of $\psi =2$ gives a morphology consistent with those typically observed. Since the outflow is ultimately driven by the entrainment of cloud material by a fast hot jet of atomic hydrogen, the outflow is in fact a layer that surrounds such gas and is unobserved at millimetre frequencies. The boundary layer then is defined by $\psi = 2.2$ for the outer edge and $\psi = 2$ for the inner edge of the outflow. These values give a cavity shape and relative thickness that appears consistent with interferometric observations of nearby low mass sources \citep[e.g.][]{lee.et.al02,arce_goodman02,jorgensen.et.al07}. 
\begin{figure}
  \includegraphics[width=80mm]{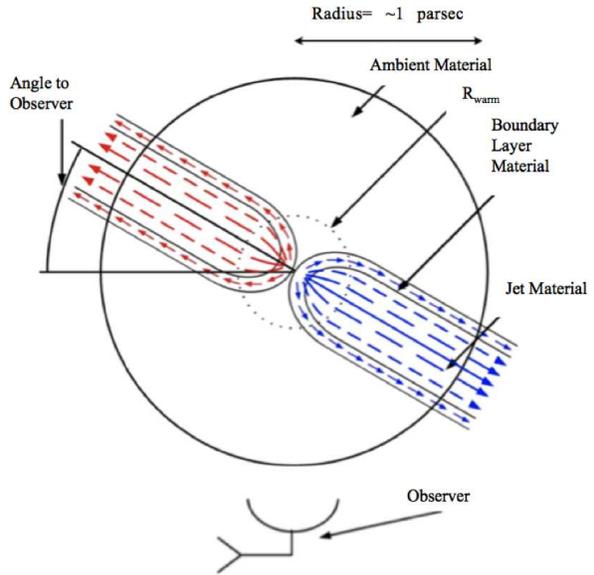}
  \caption{Sketch of the source geometry assumed for each core. The outflow cavity shape is described in the text and follows the functional form of $z = {\rm tanh}(\psi r)$ where r is the radius, $\psi$ is a fitting parameter that determines the width, and z is the height. The inner plummer sphere has radius $r = R_{\rm warm} \simeq 0.1$pc. }
  \label{fig:sketch}
\end{figure}

The orientation of the outflow to the observer must also be set and in fact this orientation is responsible for much of the variation in line profile shape seen between sources. This has been shown very clearly recently by \citep{Rawlings2013} who found that for several low-mass star forming sources, an almost identical underlying model viewed from different orientations could reproduce the complex and variety of line profiles observed. Similarly here, in preliminary modelling, we took a generic model with the outflow shape described above, and with a canonical set of parameters for massive star forming regions, and by varying the viewing angle, verified that this simple core and outflow geometry is appropriate for the three cores modelled in this work. This contributes to evidence that massive star formation proceeds in a scaled-up version of low mass star formation.
By reference to the line profile variations with viewing angle and to the GLIMPSE imagery in \citet{LoObs} we constrained the viewing angles for the sources, as detailed below in Section~\ref{sec:modelresults}.  Figure~\ref{fig:sketch} is a sketch that shows the overall source morphology that we adopt.

\subsubsection{Constrained source properties from continuum observations}
Dust continuum observations provide a measurement of the density and temperature in the core that is independent of the molecular line observations. Under the assumption that the gas and the dust are well coupled then the SED fitting analysis presented in\citet{LoObs} can be used to constrain the density and temperature in the cores. Three temperature components were required to fit the continuum observations. A `hot' gas component ($\sim 240$ K; 350 K for G333.6-0.2) is present at radii of $\sim 10^{-3}~{\rm pc}$ and will not contribute to the line emission in the Mopra beam size ($\sim 0.25~{\rm pc}$ at the distance of the G333 sources). The `warm' ($\sim 70-100~{\rm K}$) component at $\sim 0.1~{\rm pc}$ is also unresolved but may contribute to the emission from the central pointings. The bulk of the cloud is `cold' at around 20 K. (For exact values for each core see Table~\ref{tab:temp_params}.) For each source we assume that the temperature rises from cold to warm from the cloud edge to the centre according to the following form:

\begin{equation} \label{eq:T}
  T(r) = T_{\rm warm} - {\Delta\,T}\left (\frac{r}{R_{\rm warm}}\right) ^\alpha,
\end{equation} 
where the temperature range $\Delta\,T = T_{\rm warm} - T_{\rm cold}$ is derived from values from the SED fitting of Lo et al 2015 (values restated here in Table~\ref{tab:temp_params}). For the spatial scaling of the temperature variation we use $R_{\rm warm}\sim 0.1~{\rm pc}$ as also given earlier. With a weak exponent $\alpha$ of $0.2$ or $0.5$ the above form allows the temperature to change smoothly from cold to warm in the vicinity of $R_{\rm warm}$. Using a step function between the two temperatures at $R_{\rm warm}$ does not change the models appreciably and implementing a more sophisticated temperature variation would be over-interpreting the data.

Similarly, the central densities are constrained by the SED fitting. The fall off in the density away from the central peak will follow a power law such that the cloud will be observed as a distinct feature. Again, inspired from low mass star formation, we adopt a Plummer sphere profile for the density \citep{arreaga-garcia.et.al10, whitworth01}, which is a useful approximation to the more rigorous Bonner-Ebert sphere solution for a self-gravitating sphere in an environment with significant external pressure \citep{bonner56,ebert55,chandrasekhar67}. 

\begin{equation}\label{eq:rho}
  \rho(r) = \rho_0\frac{R^2_{\rm warm}} {R^2_{\rm warm} + r^2}
\end{equation} 

The Plummer profile gives an approximately constant central density which smoothly transitions into an inverse square law density drop. While other density laws can be adopted, they vary most in the central unresolved regions and would not be significantly constrained by line fitting analysis here. Since we do not resolve a significant central enhancement in the density, we again use the radius of the warm gas, $R_{\rm warm}\sim 0.1~{\rm pc}$ to set the scaling of the density law. The chosen jet and interface densities were chosen to yield outflow masses consistent with the values derived determined in Table~\ref{tab:derivedproperties}. 

\subsubsection{Constraints from line observations: turbulence} %\& chemical abundance}
From a set of line observations of different species, transitions and offsets, it is possible to readily determine the minimum line width between species from either a rare isotope species such as ${\rm H^{13}CO^+}$, by structure within a line profile shape or by a resolved hyperfine line such as HCN \citep[though see][for dealing with hyperfine line anomalies]{Loughnane2012}. The minimum line width then provides a constraint to the maximum turbulence. The minimum line width is related to the temperature and degree of turbulence in the gas
\begin{equation}
\sigma_{\rm tot}^2 = \sigma_{\rm turb}^2 + \sigma_{\rm th}^2,
\label{eqn:turb-linewidth}
\end{equation}
where $\sigma_{\rm tot}$ is the total velocity dispersion and is
calculated using $\sigma_{\rm tot}^2 = \Delta\upsilon/\sqrt{8 \ln
2}$ where $\Delta\upsilon$ is the observed FWHM of a representative 
molecular line profile. $\sigma_{\rm th}$ is the thermal velocity dispersion
given by $\sigma_{\rm T}^2 = kT/\mu$, where $\mu$ is the molecule
mass, $T$ is the gas temperature and $k$ is Boltzmann's constant. For the temperatures of the bulk of these cores ($\sim 20-30 K$) the contribution of the thermal compenent is well constrained but negligible compared to the turbulent component, $\sigma_{\rm turb}$ which dominates the line broadening. This minimum observed linewidth then provides an upperbound for the turbulence in the gas.

For the bulk of the three G333 clouds, turbulent velocities of $\sim 3-5~{\rm km~s^{-1}}$ are indicated by the line broadening. This cold highly supersonic turbulence is a characteristic feature of massive star formation that is in distinct contrast to low mass star formation where the degree of turbulence is comparable to the thermal broadening, and where the gas motions are close to subsonic. Variations in the degree of turbulence are possible and may be indicated by variations in the line width of low abundance species in high spatial resolution observations. As discussed below, we find slight evidence for increases by $\sim 1~{\rm km~s^{-1}}$ from the edge of the cloud to the central regions; this may be expected as random supersonic gas motions in the outer cloud increase along with infall motions towards the centre of the core. The following form for the turbluent velocity component is used. 

\begin{equation} \label{eq:v_turb}
  v_{\rm turb}(r) = v_{\rm turb} - \Delta\,v_{\rm turb} \left(\frac{r}{R_{\rm warm}}\right)^\beta
\end{equation}
where $v_{\rm turb}$ is the characteristic turbulent velocity (between $3.5 - 5.0~{\rm km~s^{-1}}$ measured for each source and where $\Delta\,v_{\rm turb}$ is around half the value measured for the bulk of the cloud. Again, the weak $\beta$ exponent (0.5 or 0.2) ensures that the variation in $v_{\rm turb}$ occurs smoothly in the vicinity of $R_{\rm warm}$, near the centre of the cloud. 

%move this paragraph?
%
\subsubsection{Constraints from line observations: chemical abundance}
To fix the chemical abundance, a full dark cloud chemical model should ideally be run and would be crucial across regions of changing temperature, density and desorption such as in the protostellar disk. However, again the most extreme chemical variations will be taking place at scales that are not resolved by these observations so we adopt uniform chemical abundances here for simplicity with a constant abundance for the jet and core. 

The constraints for these chemical abundances come from line observations. We initially assume a canonical abundance and compare to observed line intensities. In particular, for the modeled core masses, canonical abundances produce synthetic lines which are too strong or too self-absorbed in comparison to the observed lines. Likewise, the observed wings of the line profiles taken to be from the jets help to constrain the jet abundances and behave similarly suggesting lower abundances in the jet, generally. So, we find evidence of depletion which we discuss in Sec.~\ref{sec:depletion}.

\subsubsection{Velocity structure modelling with {\sc mollie}}
With the adoption of the source geometry, density and temperature structure, degree of turbulence and freeze-out, the cores are largely defined. To solve the equation of radiative transfer for each species, line and position requires two remaining parameters: the local velocity of the gas and the chemical abundance of the species.

The velocity structure is the key remaining quantity to be specified and along a given line of sight. The velocity of any infalling gas will vary from approximately static in the outer regions of the cloud to very large values, with a significant rotational component, as the gas approaches the disk. From the line observations (in sources where the outflow is orientated such that it does not interfere with the dynamical signature) limited constraints on the infall can be made on the basis of the degree of splitting measured in the profiles of species such as ${\rm HCO^+}$ which best trace the dense infalling gas. The separation of the line profile peaks will be due to a combination of the spread of velocities along the line of sight with the density, temperature and fraction of the beam occupied by the infalling gas. Splittings of approximately $5-10~{\rm km~s^{-1}}$ are seen, corresponding to infall velocities of a few ${\rm km~{s^{-1}}}$. We adopt the following form for the infall velocity 
\label{eq:v}
\begin{equation} 
 v(r) = v_{\rm max} \left( \frac{R^2_{\rm warm}}{R^2_{\rm warm} + r^2}\right )^\gamma
\end{equation}
%v_{\rm infall}(r) = v_0 \left(1-r/R_{\rm warm}\right)^\gamma
where $v_{\rm max}$ sets the maximum measurable infall velocity which is assumed to plateau in the central regions (when really the velocity will continue to increase on unresolvable scales) and then drop in magnitude with increasing radius. This Plummer-like scaling relation (c.f. Equation~\ref{eq:rho}) is convenient to describe this motion. Care is needed in the identification of a canonical infall velocity, to be used in semi-analytic calculations. Using the form above, we choose $v_{\rm infall}\equiv v(R_{\rm warm})=v_{\rm max}/2^{\gamma}$. With a $\gamma$ of either 1 or 2, $v_{\rm infall}$ is equal to either a half or quarter of $v_{\rm max}$.

The choice of velocity structure for the outflowing gas is more straightforward since it must be consistent with the analysis presented in Lo et al 2015. We take the outflow velocity to be uniform with values as specified in Table~\ref{tab:vel_params} and contained within the boundary layer parameterised by Equation~\ref{eq:layer}. 

Considering the possible velocity components along a given line of sight, it is possible that the line profile is assembled from an optically-thick combination of blue and red-shifted outflow gas, from a turbulent envelope and from infalling gas. A numerical solution to the radiative transfer problem is thus required. A one-dimensional code is not adequate for this because only radial motions can be modelled. Two-dimensional codes are useful \citep[e.g {\sc ratran},][]{hogerheijde_vandertak2000} because many systems will have cylindrical symmetry around the outflow axis. In fact, because of optical depth effects, the line profile  changes dramatically as the angle varies between the outflow axis and the observers 
line of sight \citep{ward_thompson_buckley01,Rawlings2013}. To allow for arbitrary viewing angle relative to the symmetry axis it is therefore necessary to couple even a 2-D geometry to 3-D ray-tracing algorithm. 

The 3-D molecular line radiative transfer code used throughout this work was written and developed by Keto and collaborators \citep[see][for examples of its use]{Keto2004,Rawlings2004,Redman2006,Carolan2008,Lo2011}. The code, {\sc mollie} (MOLecular LIne Explorer), is used to generate synthetic line profiles to compare with observed molecular rotational transition lines. {\sc mollie} splits the overall structure of a cloud into a 3-D grid of distinct cells where density, abundance, temperature, velocity and turbulent velocity are defined as described above. In order to calculate the level populations, the statistical equilibrium equations are solved using an Accelerated Lambda Iteration (ALI) algorithm \citep*{Rybicki1991} that reduces the radiative transfer equations to a series of linear problems that are solved quickly even in optically thick conditions. For an arbitrary viewing angle to the model cube, ray-tracing is then used to calculate the molecular line intensity as a function of velocity, for a set of positions that matches in number the fixed gridding of the model. 

{\sc mollie} has been benchmarked against a suite of problems \citep{van_Zadelhoff2002} and all the models were found to reproduce the test observations to within a few per cent. We emphasise that this benchmarking plus the model description presented here, means it should be possible to replicate our model line profiles with other 3-D line radiative transfer codes \citep[e.g. {\sc lime},][]{brinch_hogerheijde2010} and that such line radiative transfer codes have been used successfully since the mid 1990s  \cite{Rawlings1992,Choi95,hogerheijde_vandertak2000,ward_thompson_buckley01,van_Zadelhoff2002}.

For the modelling in this paper for example, we used a model cube with an edge of 2  pc with $16 \times 16 \times 16$ cells to output a $16 \times 16$ grid of line profiles. The profiles from a central subset of the model were averaged to match the Mopra beam, to compare with the observed data of the central spectra in Figure~\ref{fig:profiles}. The models were found to converge within 10 iterations so that longer runs were unnecessary. The calculated rays are reduced by {\sc mollie} from the initial requirements of the set up parameters of the model to a subset suited to the resolution of the model; for instance, it discards some rays as redundant or not crossing the model. For this modelling, approximately 3000 rays is reduced to approximately 2500 and divided up across views down to $\approx 200$ with some variance across molecules and cores, and a requested 401 channels is smoothed to 200 channels.

We choose a density which yields a mass in agreement with the observations. Likewise, we set the temperatures of the model per the observational SED fits. Then we find a power for the temperature law that gives a good match in intensity to the most observed lines from the core, where possible (variance across the cores makes this choice difficult, see Figures~\ref{fig:modelprofiles02}, \ref{fig:modelprofiles04}, \& \ref{fig:modelprofiles05}). Using the minimum observed linewidth as a guide to the maximum possible contribution from turbulent broadening, we select $v_{\rm turb}$. We next fit the velocity structure such that features such as infall are reproduced. The exponents then for velocity and turbulence are again chosen to agree with both the grids and retaining velocity features in the profiles. Lastly, we adjust discrepancies in intensity with our choice of chemical abundance. This process is then iterated as needed prioritizing minor changes in the least constrained parameters being (in increasing order of constraint) abundance, exponents for power laws (though their primary constraint comes in the form of the grids of data), turbulence, velocity, density, and temperature.

\subsubsection{Determining properties and uncertainties from the best fit models}
Following the method given above, individual models (discussed below) were made for each of the three cores and for each source slight differences in the exponents of the power laws for turbulence and temperature were adopted as well as variations in the abundance. The power law exponents used can be found with the other modelled values in Tables \ref{tab:vel_params} and \ref{tab:temp_params}. An exploration of the exactness of the model fit was carried out for each parameter to estimate the uncertainty in the best fit model. The quoted range fits the data to within 20\%. These errors could be used to estimate the uncertainty on subsequent derived quantities but for clarity we use the adopted best fit values in the analyses presented below. %are collected together in Table~\ref{tab:powerparams}.

{\if powers}
\begin{table}
  \caption {Power law exponents used in the modeling: (1) source ID, (2) the exponent of the power law for temperature, (3) the exponent of the power law for turblent velocity, (4) size of change of the turbulent velocity, and (5) the exponent of the power law for velocity.}
  \label{tab:powerparams}
  \centering

 \label{tab:powerparams}
  \centering
  \begin{tabular}{@{}ccccc}
    \hline
    ID & $\alpha$ & $\beta$ &  $\Delta\,v_{\rm turb}$ & $\gamma$ \\
     &  &  & \kms &  \\
    (1) & (2) & (3) & (4) & (5) \\
    \hline
    G333.6--0.2 &0.2 & 0.33& 1.6 & 3 \\
    G333.1--0.4 & 0.5 & 0.2 & 1.5 & 2 \\
    G332.8--0.5& 0.25 & 0.33 & 2.5 & 1\\
    \hline
  \end{tabular}
\end{table}
\fi

The modelling does not suggest that these power laws are absolute but rather that model found is an exemplar from a family of models with similar power laws and constants, each adjusted by underlying rules to arrive at another member of the family of models. For instance, since density, temperature, and abundance all contribute to the intensity of the line profile, an increase or decrease in one requires a decrease or increase in the others. Certain molecules and lines (e.g. \ce{CS}(2--1) and \ce{CO}(1--0)) are more sensitive to changes in temperature and density and so offer further constraints on the free parameters of these power laws.

The model results are quite robust in that, as argued above, many of the parameters are constrained in advance of the line profile radiative transfer model calculation. Based on repeated modelling of these sources over variations in those parameters that are not already tightly constrained by the observations, we report estimates for the accuracy of the model results. Example chi-squared calculations were carried out as a more quantitative check on these estimates but it is not practical to do this formally over all possible combinations of parameters for all 18 modelled line profiles. We give uncertainties that err on the large side to allow for this shortcoming. Finally, we give a table of the percentage difference between the integrated line intensities and the best model fit that are consistent with our model uncertainties. We find overall deviations of typically $15-20\%$ overall, a figure which does not quite capture the degree to which the individual line profile features are captured. Given the limitations to the modelling procedure which include a simple spherical geometry, no chemical or hydrodynamic model and intrinsic uncertainties in the molecular collision data, the fitting appears satisfactory.

\subsection{Modelling results} \label{sec:modelresults}
\begin{figure*}
  \includegraphics[height=0.95\textheight]{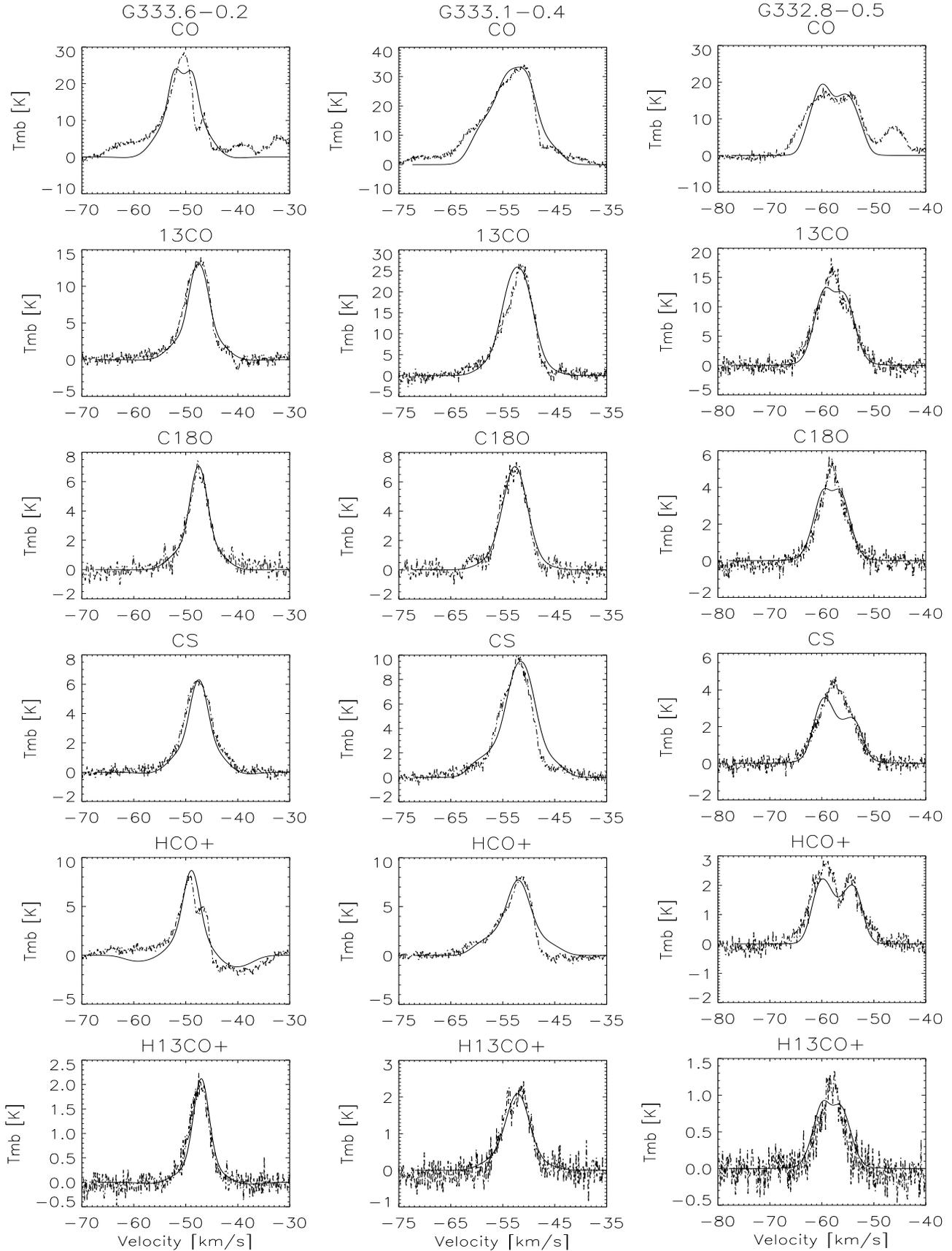}
  \caption{Radiative transfer line models (solid) overlaid on the observed spectra (dashed) for each of the sources. From top to bottom: CO, \ce{^{13}CO}, \ce{C^{18}O}, \ce{CS}, \ce{HCO+} and \ce{H^{13}CO+}. Left column: G333.6--0.2, middle column: G333.1--0.4, right column: G332.8--0.5. Intensity is in main beam brightness temperature.}
  \label{fig:modelprofiles}
\end{figure*}

\begin{figure*}
  \includegraphics[width=160mm]{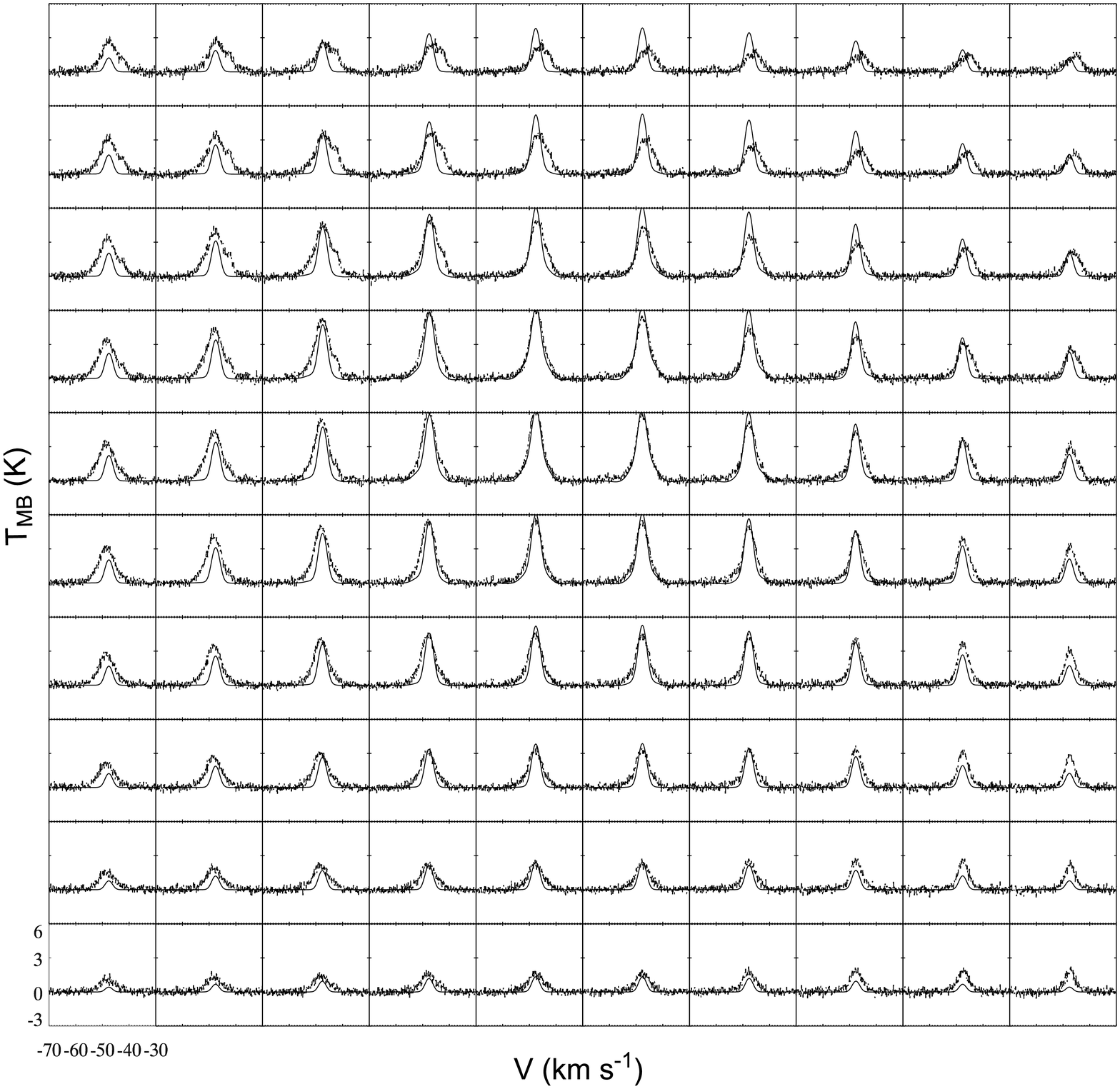}
  \caption{Modelled \ce{CS} line (solid) overlay on a grid of observed emissions (dashed) of G333.6--0.2. Intensity is in brightness temperature, the grid is centred at RA=16:22:07, Dec=-50:06:21, spacing between each of the spectra is 12 arcseconds.}
  \label{fig:modelprofiles02}
\end{figure*}

\begin{figure*}
  \includegraphics[width=160mm]{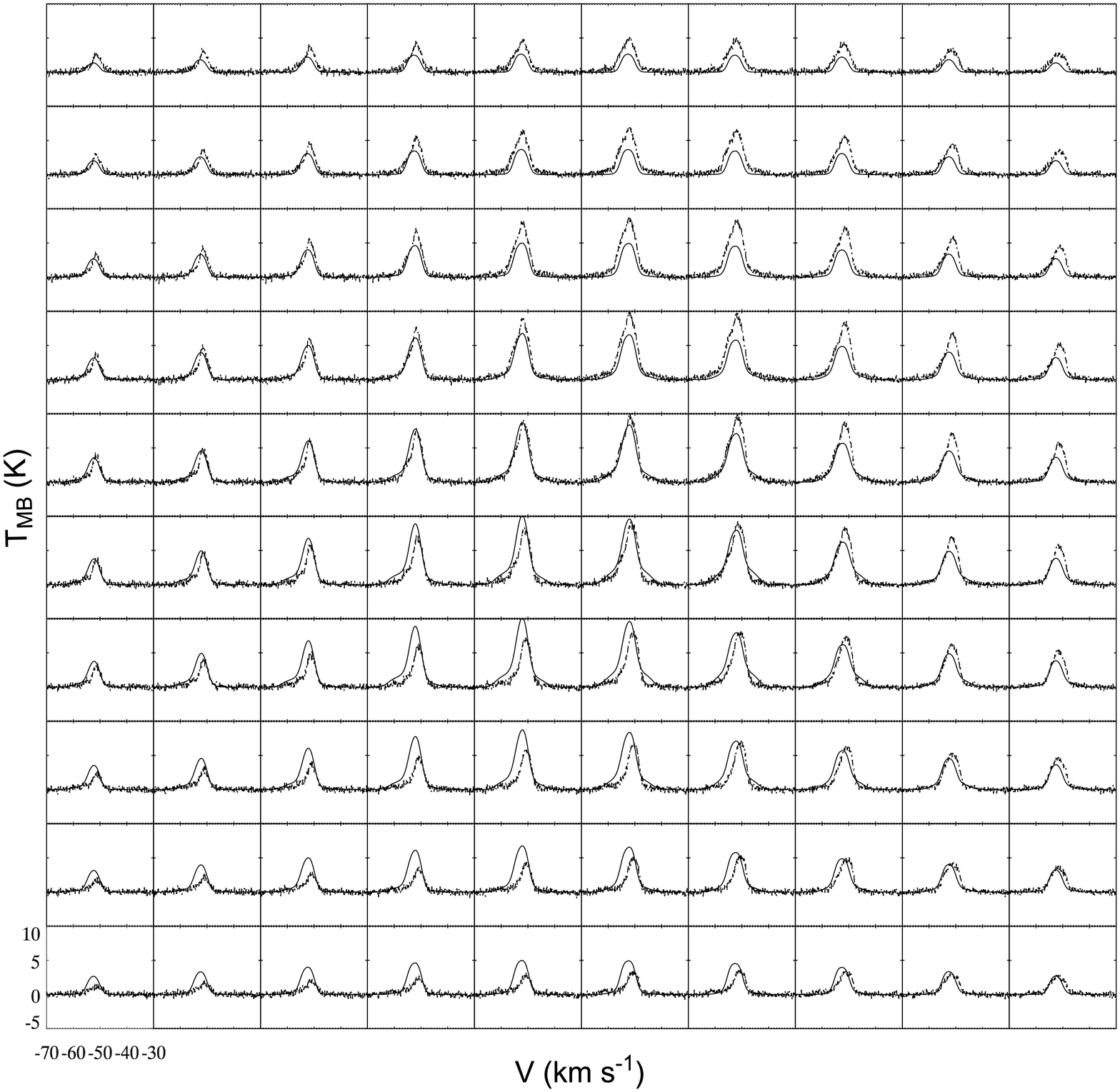}
  \caption{Modelled \ce{CS} line (solid) overlay on a grid of observed emissions (dashed) of G333.1--0.4. Intensity is in brightness temperature, the grid is centred at RA=16:21:03, Dec=-50:35:12, the spacing between each of the spectra is 12 arcseconds.}
  \label{fig:modelprofiles04}
\end{figure*}

\begin{figure*}
  \includegraphics[width=160mm]{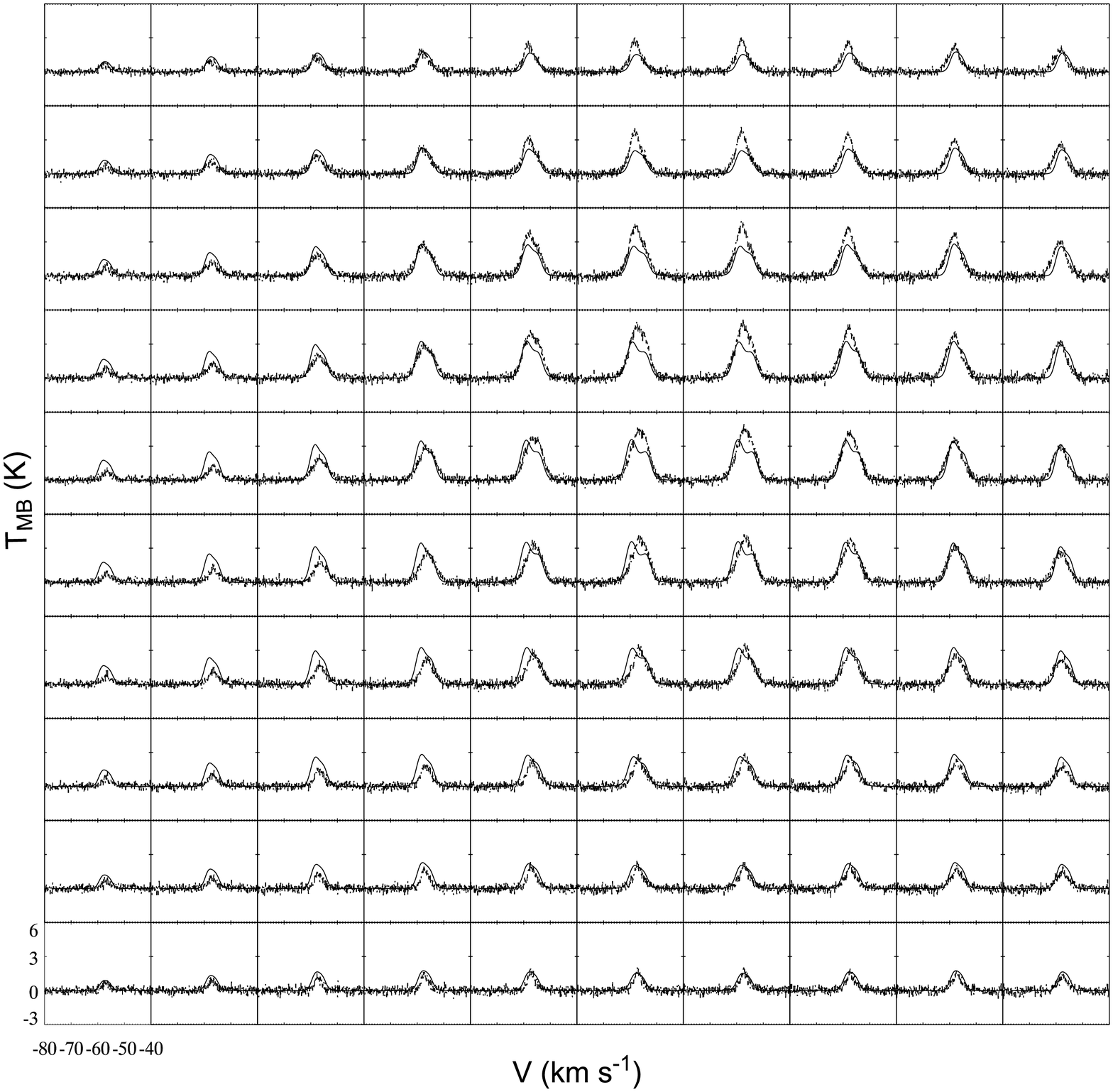}
  \caption{Modelled \ce{CS} line (solid) overlay on a grid of observed emissions (dashed) of G332.8--0.5. Intensity is in brightness temperature, the grid is centred at RA=16:20:10, Dec=-50:53:14, the spacing between each of the spectra is 12 arcseconds.}
  \label{fig:modelprofiles05}
\end{figure*}

In the following paragraphs, we discuss each of the three outflow sources in turn, highlighting particular features or additions to the general modelling procedure described above.
The modelled synthetic line profiles of the central profile of the three sources are shown in Figure \ref{fig:modelprofiles} along with the modelled results listed in Tables \ref{tab:vel_params}, \ref{tab:temp_params} and \ref{tab:model_abund}. A grid of modelled spectra of \ce{CS} emission of the three sources are shown in Figures \ref{fig:modelprofiles02}-\ref{fig:modelprofiles05}. Similar grids of profiles were produced for the other lines to verify that the model results gave satisfactory fits to the line profiles from across the clouds.

\begin{table*}
  \caption{Modelled velocity parameters of the three sources, the columns are as follows: (1) source name; (2) the central infall velocity at $R_{\rm warm}$; (3) velocity exponent in equation 5; (4) outflow velocities; (5) peak core turbulent velocity; (6) drop in core turbulent velocity; (7) exponent of the turbulence in equation 5; (8) outflow turbulent velocities.}
  \label{tab:vel_params}
  \begin{tabular}{@{}cccccccc}
    \hline
Source & $v_{\rm infall}$ & Velocity & $v_{\rm outflow}$ & $v_{\rm turb, core}$ & $\Delta v_{\rm turb, core}$\footnotemark & Turbulence & $v_{\rm turb, outflow}$ \\
 & \kms & exponent $\gamma$ & \kms & \kms & \kms & exponent $\beta$& \kms \\
(1) & (2) & (3) & (4) & (5) & (6) & (7) & (8) \\
 \hline
G333.6--0.2 & 2.4$\pm$0.5 & 2.0$\pm$0.5 & 9.5$\pm$0.5 & 3.5$\pm$0.5 & 2.0$\mp$0.5 & 0.5$\pm$0.2 & 3.5$\pm$0.5 \\
G333.1--0.4 & 1.6$\pm$0.25 & 2.0$\pm$0.5 & 6.5$\pm$0.5 & 3.5$\pm$0.5 & 1.5$\mp$0.5 & 0.2$\pm$0.1 & 3.5$\pm$0.5 \\
G332.8--0.5 & 3.5$\pm$0.5 & 1.0$\pm$0.5 & 7.0$\pm$0.5 & 3.8$\pm$0.5 & 2.0$\mp$0.5 & 0.5$\pm$0.2 & 3.8$\pm$0.5 \\
\hline
  \end{tabular}
  \\ $^1$\footnotesize{$\Delta v_{\rm turb, core}$ is anti-correlated to increasing the other turbulence parameters; hence the use $\mp$ rather than $\pm$.}
\end{table*}
\begin{table*}
  \caption{Modelled temperature and density parameters of the three sources: (1) source name; (2) $\phi$ viewing angle relative to the observer; (3) warm gas temperature; (4) temperature exponent in equation 2; (5) cold gas temperature; (6) outflow temperature; (7) peak core density; (8) outflow density; (9) derived kinetic infall mass rate.}
  \label{tab:temp_params}
  \begin{tabular}{@{}ccccccccc}
\hline
Source & $\phi$ & $T_{\rm warm}$ & Temperature & $T_{\rm cold}$  & $T_{\rm outflow}$ & $n_{0}$ & $n_{\rm outflow}$ & $\dot{M}_{\rm infall}$ \\
             & $^\circ$ & K & exponent $\alpha$ & K & K &  $10^6$\,cm$^{-3}$ &  $10^6$\,cm$^{-3}$ & $10^{-3}$~\solarM\,yr$^{-1}$ \\
(1) & (2) & (3) & (4) & (5) &  (6) & (7) & (8) & (9) \\
\hline
G333.6--0.2 & 70$\pm$10 & 98 & 0.5$\pm$0.2 & 28$\pm$5 & 23$\pm$5 & 5.7 & 3.3 & 23 \\
G333.1--0.4 & 60$\pm$5 & 69 & 0.5$\pm$0.2 & 19$\pm$5 & 30$\pm$5 & 3.5 & 0.50 & 9 \\
G332.8--0.5 & 0$\pm$5 & 68 & 0.2$\pm$0.1 & 20$\pm$5 & 20$\pm$5 & 2.2 & 1.2 & 12 \\
\hline
  \end{tabular}
\end{table*}

\subsubsection{G333.6--0.2}
The observed line profiles of G333.6--0.2 are heavily affected by the nearby source of continuum radiation; \ce{^{12}CO}, \ce{^{13}CO} and \ce{HCO+} (Figure~\ref{fig:profiles}) clearly show deep red-shifted continuum absorption. In radiative transfer models, the background continuum radiation field is usually set to be the cosmic microwave background and so absorption features are not expected to be seen. The presence of a nearby powerful source of continuum radiation was accounted for in the model by adopting the method of \citet{Redman2003} in which the core is bathed in a grey-body radiation field, here of T$_{\rm grey}$=24K. There is an absorption effect noted in the \ce{^{13}CO} and \ce{CO} spectra extending from $-50$~\kms to $-30$~\kms corresponding roughly to the absorption effect in \ce{HCO+} suggests that the grey-body field may not well model the effects of this H{\sc ii} region on these molecules, specifically \ce{CO}. In particular, in \ce{CO}, a number of secondary peaks are noted in the red wing of \ce{CO}; these peaks are localized to the right side of the observed field maps. The strong sharp peak from $-50$ to $-45$~\kms is present in only the lower right quadrant. The furthest removed peaks in the blue and red ($\sim -75$~\kms and $\sim-30$~\kms) are only present in the centre right of the field (where they peak strongly). This suggests that all of these effects are from contaminating clouds along the line of sight.

The \ce{HCO+} profile is an inverse P-Cygni profile with absorption in the red rather than blue wing. The suggestion from the modeling is that the colder jet is seen in absorption against the greybody due to the H{\sc ii} region. This strongly constrains T$_{\rm grey}$, T$_{\rm jet}$, and T$_{\rm interface}$ to have a ladder of temperatures with T$_{\rm grey}$ between that of T$_{\rm interface}$ above and T$_{\rm jet}$ below. The channel maps (see Figure~\ref{fig:modelprofiles04}) further suggest that the greybody is not isotropic; however, implementation of such a localized field in {\sc mollie} would be a considerable extension to the code and would introduce a number of free parameters to constrain. 

%The \ce{HCO+} profile is an inverse P-Cygni profile with absorption in the red rather than blue wing. The suggestion from the modeling is that the colder jet is seen in absorption against the greybody due to the H{\sc ii} region. This strongly constrains T$_{\rm grey}$, T$_{\rm jet}$, and T$_{\rm interface} to have a ladder of temperatures with T$_{\rm grey}$ between that of T$_{\rm interface}$ above and T$_{\rm jet}$ below. The channel maps (see Figure~\ref{fig:modelprofiles04}) further suggest that the greybody is not isotropic; however, implementation of such a localized field in {\sc mollie} would be a considerable extension to the code and would introduce a number of free parameters to constrain. %This profile was accomplished by adding a temperature gradient falling from $1.25 \times T_{\rm warm}$ to $T_{\rm warm}$ to the inner Plummer sphere, and adding a strong stellar wind to the core as an expanding envelope around the source, the expected mechanism underlying P-Cygni profiles \citep{Thompson2009}. The inner Plummer sphere was increased from 0.1~pc to 0.125~pc in order to better resolve the effects of these changes. This model reproduces the shape of the \ce{HCO+} profile. 

The low intensity wings of extended emission in \ce{CO} and \ce{HCO+} are sensitive to changes in the jet and interface temperatures and abundances as well as the greybody temperature: in particular, that the red wing appears in absorption while the blue wing appears in emission might be captured by giving the interface a higher temperature or density than the jet. This is effectively the interface screening for the absorption of the jet against the greybody but this effect not fully explored and captured in the present model.

The other molecules however are not particularly sensitive to the choice of these three parameters. The absence of an absorption feature in the \ce{N_2H^+} profile shown in \citet{LoObs} is consistent with the general understanding that this tracer is never seen in outflows, as \ce{N_2H^+} is destroyed by reaction with \ce{CO} \citep{Tafalla2004, Lo2009}. While for \ce{CS}, with its higher critical density than \ce{CO} and not particularly temperature sensitive, the absence of absorption feature is expected. This is further supported by the existence of inverse P-Cygni absorption feature in \ce{HNC} profile, as it is shown that \ce{HNC} is more sensitive to low temperature \citep[e.g.][]{Hirota1998}, which is also seen throughout the G333 cloud \citep{Lo2009} and dense cold core \citep{Lo2007}. Furthermore, the absorption feature in \ce{HCN} profile is stronger at the higher velocity end of outflow wing, \ce{HNC} only shows absorption at the lower velocity end and is much weaker, while the \ce{CS} outflow profile is `smooth' and does not show signs of absorption.

\subsubsection{G333.1--0.4}
The observed \ce{CO} and \ce{HCO+} line profiles show red-shifted self-absorption and splitting characteristic of infall over the core. However, the lack of distinctive absorption features toward the center of the cloud is due to the orientation of the cloud: the blue shifted outflow is projected against the centre of the core which tends to obscure the infall signature (indeed, inspection of the same model rotated to a viewing angle of 0 degrees shows a line profile exhibiting a clear double peak infall signature very similar to those of G332.8--0.5).The synthetic line profiles show some evidence of being affected by continuum radiation (the red side of the double-peaked profile infall is suppressed almost into absorption), likely provided by the H{\sc ii} region. The {\it Spitzer} GLIMPSE image in \citet{LoObs} shows the location of the embedded H{\sc ii} region with respect to the molecular emission from G333.1--0.4. The abundances were increased so that line intensities matched those observed converted to $T_{\rm MB}$, and interface and outflow abundances were then balanced to give the lines their distinctive raised shoulders in the red wings.

\subsubsection{G332.8--0.5}
In G332.8--0.5, the observed \ce{HCO+} line is clearly split with a stronger blue-shifted peak (the corresponding optically thinner tracer, \ce{H^{13}CO+} displays a single peak at the \ce{HCO+} line centre, eliminating multiplicity as the origin of the two peaks in \ce{HCO+}). The embedded H{\sc ii} region in this source is much smaller and less powerful than those in the other two. There is the suggestion of a red-shifted interloping clump in the triply-peaked \ce{^{12}CO} line profile, although inspection of the channel maps suggests that this is just structure in the outflow wing.  In the individual model for this core, the relative intensities of \ce{^{12}CO} and \ce{HCO+} were balanced between sufficiently higher abundances and velocities than in the generic model such that their absorption features were reproduced. The synthetic profiles for this source provide a reasonable match to those observed when a moderate depletion is applied.  A factor of 0.1 is applied for $r < 0.5~{\rm pc}$ for all lines other than \ce{CO} and \ce{HCO+}. \ce{CO} is depleted by a factor of 0.5 rather than 0.1, and \ce{HCO+} is not depleted. Without this depletion effect, the lines other than \ce{CO} and \ce{HCO+} display a weak infall feature similar to \ce{CO} and \ce{HCO+}. The slight absorption features in the modelled \ce{CS} but not in the observations suggest that the velocity found to balance \ce{^{12}CO} and \ce{HCO+} is overly large; however, it cannot be much reduced and retain the pronounced infall features in \ce{^{12}CO} and \ce{HCO+}. A second interpretation might be that the second peak in \ce{CS} is physical but is absorbed against the background, implying that there is some absorption in the blue-shifted wing. As this would introduce an absorption factor similar to G333.6--0.2, we have not attempted to model it as it introduces more free parameters, and as given the relative strength of the two H{\sc ii} regions, it is doubtful the greybody approximation used in G333.6--0.2 would be a reasonable approximation here. 

\subsection*{}
The percent difference between modelled and observed signal, averaged over the channels in which the profile is present, for each core is as follows: G333.6--0.2 7.8\%, G333.1--0.4 9.7\%, and G332.8--0.5 13\%. By itself, the percentage difference in flux is not a helpful quantitative measure of the accuracy of a model (as of course a ``perfect'' match could be created with a simple gaussian of the right dimensions which failed to capture any of the features of the observation at all); however, it provides a quantitative measure for the models' agreement alongside such visual, qualitative results as Fig.~\ref{fig:modelprofiles} and Figures~\ref{fig:modelprofiles02},\ref{fig:modelprofiles04}, \&~\ref{fig:modelprofiles05}. For CO in G333.6--0.2 and G332.8--0.5, we integrate over the signal excluding the structures in the blue wing in \ce{^{12}CO} as these were intentionally unmodelled. We emphasize that collisional rates are known to only $\sim$ 10\% and that as our models are near this percent uncertainty further work would be an over-interpretation of the data.

\subsection{Mass infall rate and timescales} \label{sec:infall}
The modelled infall velocity of two of the sources are around $\sim 3$ \kms\ (Table \ref{tab:vel_params}). Turbulence and infall contribute to the line width, shoulders and infall signature of the abundant \ce{CO} and \ce{HCO+} line profiles. For a much lower infall velocity, structures such as the shoulder and the infall signature in the line profiles will be masked out by turbulence. For G332.8--0.5, the double peak infall feature in \ce{CO} and \ce{HCO+} line profiles is delicately balanced on the turbulent velocity (i.e the line broadening). While the `lack of infall signature' in the G333.1--0.4 line profile is due to the viewing angle in which the outflow masks the expected double-peaked infall signature, an infall component remains necessary however for consistency. 

Using the model results we estimate the kinematic infall mass rate as outlined in \citet{Myers1996},

\begin{equation}\label{eq:m_in}
  \dot{M}_{\rm infall} = 4 {\rm \pi} r_{\rm infall}^2 n\mu m_{\rm H_2} v_{\rm infall},
\end{equation}

\noindent where $\mu = 1.2$ as before, $n$ is the density obtained from the models and the infall radius $r_{\rm infall}$ is taken to be the warm component in the SED fitting (Lo et al 2015), which is also $R_{\rm warm}=0.1~{\rm pc}$ in the models. This rough mass infall rate estimate leads to values of $~10^{-2}$ \solarM\,yr$^{-1}$ (Table \ref{tab:temp_params}) which are a factor of a few higher than the mass outflow rates of $~10^{-3}$ \solarM\,yr$^{-1}$ (given in Table~\ref{tab:derivedproperties}). Here we have recalculated the infall using $r_{warm}$ rather than the full radius of the core which was used in \citet{LoObs}; we do this to reflect the modelled variable infall velocity and give a lower limit of the infall time.

The simplest estimate of the collapse timescale is just the dynamical timescale of $t_{\rm dyn}\sim r_{\rm infall}/v_{\rm infall}$. However the gas will accelerate as collapse proceeds so one alternative timescale is one in which the gravity of the mass enclosed at the infall radius is taken into account: 
\begin{equation}\label{eq:t_infall}
  t_{\rm infall} = \frac{M_{\rm infall}}{\dot{M}_{\rm infall}} \simeq \frac{M_{\rm dust}}{\dot{M}_{\rm infall}}
\end{equation} 
Finally, the free-fall timescale in a self-similar isothermal sphere, used in calculations of relatively quiescent cores in low-mass star-forming regions, can also be calculated 
\begin{equation} \label{eq:t_ff}
  t_{\rm ff} =  \sqrt\frac{3 \pi}{ 32 G \rho}.
\end{equation}
We likewise recalculate the freefall time from $r_{warm}$ as well so that it is comparable to our calculated infall times. These different timescales of the three sources are listed in Table \ref{tab:timescale} along with the outflow timescale from Lo et al (2015), which is not changed between the models and observations.  It can be seen that that the infall velocity implies that the cores are collapsing somewhat slower than the free-fall timescale. All the timescales are of order $10^4~{\rm yrs}$ and assuming a constant infall rate, we estimate the core's lifetime by dividing the core mass (reported in Table~\ref{tab:derivedproperties}, and consistent with the modelled core density) with the infall rate, giving a lifetime of $\sim 10^5$ years. 

\begin{table}
  \centering
  \caption{Estimated time-scales of the three cores: (2) dynamical timescale, (3) infall timescale from Equation~\ref{eq:t_infall} (4) free-fall timescale Equation~\ref{eq:t_ff} and outflow timescale from Table \ref{tab:derivedproperties}.}
  \label{tab:timescale}
  \begin{tabular}{@{}ccccc}
    \hline
    Source & $t_{\rm dyn}$ & $t_{\rm infall}$ & $t_{\rm ff}$  & $t_{\rm outflow}$ \\
     & $10^4$ yrs & $10^4$ yrs & $10^4$ yrs & $10^4$ yrs \\
    (1) & (2) & (3) & (4) & (5) \\
    \hline
    G333.6--0.2 & 3.9 & 7.3 & 5.6 & 1.0 \\
    G333.1--0.4 & 5.9 & 10 & 4.3 & 4.0 \\
    G332.8--0.5 & 2.7 & 4.4 & 4.2 & 8.0 \\
    \hline
  \end{tabular}\\
\end{table}

\begin{table}
  \caption{Individual modelled abundances. Columns are as follows: (1) Source, (2) molecule (with source given as a heading); (3) and (4) the envelope and outflow abundances. These abundances are not corrected for freeze-out and so may be lower than expected typical canonical abundances. The CO abundance is poorly constrained due to very high optical depth and source confusion and so is not reported here.}
  \label{tab:model_abund}
  \centering
  \begin{tabular}{@{}cccc}
    \hline
    Source & Molecule & $\chi_{\rm envelope}$ & $\chi_{\rm outflow}$ \\
     & & $n_{\rm H_2}^{-1}$  & $n_{\rm H_2}^{-1}$  \\
    (1) & (2) & (3) & (4) \\
    \hline
%    G3333.6-0.2 &  &  &  &  &  & \\
%    G3333.6-0.2 & \ce{CO} & $3 \times 10^{-6}$ & $3\times 10^{-7}$ & 0.8\\
      G3333.6-0.2   & \ce{^{13}CO} & $3 \times 10^{-7}$ & $3\times 10^{-8}$  \\
     & \ce{C^{18}O} & $1 \times 10^{-7}$ & $1 \times 10^{-8}$ \\
     & \ce{CS} & $3\times 10^{-10}$ & $5 \times 10^{-10}$  \\
     & \ce{HCO+} & $1\times 10^{-10}$ & $4\times 10^{-10}$  \\
     & \ce{H^{13}CO+} & $2\times 10^{-11}$ & $2\times 10^{-11}$  \\
%    G333.1-0.4 &  &  &  &  &  & \\
 %   G333.1-0.4 & \ce{CO} & $2 \times 10^{-6}$ & $7 \times 10^{-6}$  & 0.6\\
 G333.1-0.4      & \ce{^{13}CO} & $7 \times 10^{-7}$ & $7 \times 10^{-7}$  \\
     & \ce{C^{18}O} & $1 \times 10^{-7}$ & $1 \times 10^{-7}$  \\
     & \ce{CS} & $5 \times 10^{-10}$ & $5\times 10^{-9}$ \\
     & \ce{HCO+} & $8\times 10^{-11}$ & $8 \times 10^{-10}$ \\
     & \ce{H^{13}CO+} & $2 \times 10^{-11}$ & $6 \times 10^{-11}$  \\
%    G332.8-0.5 &  &  &  &  &  & \\
%    G332.8-0.5 & \ce{CO} & $7 \times 10^{-7}$ & $7\times 10^{-8}$ & 0.95 \\
G332.8-0.5     & \ce{^{13}CO} & $1 \times 10^{-7}$ & $1 \times 10^{-7}$ \\
     & \ce{C^{18}O} & $3 \times 10^{-8}$ & $3 \times 10^{-8}$  \\
     & \ce{CS} & $7 \times 10^{-10}$ & $1 \times 10^{-10}$ \\
     & \ce{HCO+} & $4 \times 10^{-11}$ & $4 \times 10^{-12}$  \\
     & \ce{H^{13}CO+} & $5 \times 10^{-12}$ & $2 \times 10^{-12}$ \\
    \hline
  \end{tabular}
\end{table}

\section{DISCUSSION} \label{sec:discussion}
\subsection{Dynamics of the cores}
The peak infall velocities of the three cores are high but are not in discord with the turbulent velocity widths and the outflow velocities (Table \ref{tab:vel_params}), and indeed from general arguments based on the virial theorem. In this way the cores behave as scaled-up versions of low mass star forming cores. However, because the temperatures are low all the gas motions are highly supersonic, in contrast to low mass star formation where the motions are approximately sonic for turbulence, marginally supersonic for infall and supersonic for outflows. 

The estimated infall timescales of the cores are comparable to but slightly longer than the free-fall timescale. This is consistent with some internal form of pressure, such as turbulence, that is partially supporting the cloud against collapse. In some situations, the high dynamical activity in high-mass star forming regions can enhance the rate of collapse of cores \cite{Lintott2005}. This effect does not seem to be present in our cores, though interestingly, \citet{Lintott2005} show the approach of assuming an accelerated collapse in chemical models explains the observed difference in behaviour of tracer species such as \ce{N_2H^+} and \ce{CS} between high- and low-mass star formation. In fact, for  G3333.6-0.2 and G332.8-0.5, \citet{LoObs} do observe that \ce{CS} has a similar distribution to 1.2-mm dust emission while the \ce{N_2H^+} emission is offset \citep{Lo2009} in the way described in \citet{Lintott2005}. 

\subsection{Depletion due to freeze out} \label{sec:depletion}
Molecular line emission can be strongly affected by freeze-out of molecules onto the surface of dust grains, which obviously prevents them from emitting rotational line radiation, with the CO isotopologues particularly affected. Estimates of the total gas column density implied by an optically thin molecular line and by the dust continuum can be used to quantify the degree of depletion. This can also be readily seen in radiative transfer modelling where depletion is required to avoid line profile strengths that are far too large for affected species. Freeze out proceeds fastest in the central denser warmer regions of the cloud (until the sublimation temperature is reached at which point the grain mantles are re-released back into the gas phase in the hot core phase of massive star formation) and this is modelled by adopting a characteristic freeze-out radius, inside which the gas is significantly depleted out of the gas phase.

Using a canonical CO abundance of $1.5\times 10^{-5}$ with respect to molecular hydrogen would give a gas mass for the cores that matches the dust mass given in Table~\ref{tab:derivedproperties}. However, such an abundance gives line profiles shapes that are completely inconsistent with the observations, with heavy self-absorption and saturation. The most likely explanation is that at the temperatures of the outer regions of the cores, molecules such as CO will readily freeze onto dust grains. This is consistent with the strong detection of N2H+ for these cores \citep{LoObs} since this molecule is anticorrelated with CO (as spectacularly shown  by \citet{qi13}). In the central regions, where $T>T_{\rm warm}$, the molecules will resist freeze-out or the ice mantles will desorb, and indeed the rich chemistry of the later hot core phase is driven by the release of ices back into the gas phase. The different CO isotopologues probe the freeze out in different ways. Because the $^{12}$CO is so optically thick, the central regions of the core are not detected whereas the two rarer isotopologues will be able to detect the higher gas phase abundance in the inner warm region. Therefore, the typical derived abundances for the cores reported in Table~\ref{tab:model_abund} include a degree of freeze-out. We estimate that for G333.6-0.2, the degree of depletion is 0.8; for G333.1-0.4, it is 0.6 while for G332.8-0.5 a high value of 0.95 is indicated. These values are consistent with those detected in nearby low mass star forming cores, where CO depletions of over 90\% can be seen \citep{christie12}.

\subsection{Fractionation of \ce{^{13}C}-bearing species} \label{sec:enhancement}
Even accounting for the effects of freeze-out, the modelling suggests a best match to the observed line profiles with a \ce{^{12}CO}:\ce{^{13}CO} ratio of around 10, which is significantly lower than the value ($\sim 50$) predicted by the isotopic abundance gradients in the Galaxy \citep[e.g.][]{Langer1990,Milam2005} for a source at a galactocentric radius of 5.5 kpc. We inspected all the molecular line data for which we had observed both the \ce{^{12}C} and \ce{^{13}C} isotopes, which included the two CO isotopologues in addition to those of \ce{HCO+} and \ce{HCN}. The \ce{H^{13}CO+} emission is particularly bright towards the three outflow sources, with a \ce{HCO+}:\ce{H^{13}CO+} line ratio of $\sim 4$, compared with $\sim 12$ throughout the rest of the GMC. Similarly, the \ce{HCN}:\ce{H^{13}CN} ratio towards the outflow sources is $\sim 6$ whereas in the rest of the GMC it is typically $\sim 15$. In addition, the \ce{H^{13}CO^+} emission from G333.6--0.2 (Figure~\ref{fig:modelprofiles}) is slightly self-absorbed towards the line centre, and \ce{HCO+} and \ce{^{12}CO} are both optically thick for this source. Taken together across the molecular species, we find a \ce{^{12}C}:\ce{^{13}C} ratio of $\sim 6$. The poorly constrained \ce{^{12}CO} abundance and the optical thickness of the CO and \ce{HCO+} lines suggest that these ratios are only lower limits.

These enhanced ratios suggest isotopic fractionation effects are taking place in the sources. In particular, \ce{^{13}CO} is known to be enhanced relative to \ce{^{12}CO} at $T_{\rm ex} \la 35\, K$, \citep{Mladenovic2014} this is due to the exothermic reversible reaction through which it is produced,
\begin{equation} 
  {\rm ^{13}C^+ + ^{12}CO \rightarrow ^{12}CO^+ + ^{13}CO } + \Delta E
\end{equation}
\noindent where $\Delta E = 35$\,K. For temperatures $T_{\rm ex} \la 35$\,K, the reversed endothermic reaction is suppressed. The presence of \ce{N2H+} in all three sources \citep{LoObs} is consistent with such low temperatures. \citet{Mladenovic2014} carry out detailed theoretical calculations of the basic isotope exchange reactions involved in the $^{12}$C/$^{13}$C and $^{16}$O/$^{18}$O balance. In particular reactions between the commonest C$^+$, CO, and HCO$^+$ isotopologues were carried out. From these reactions, the degree of isotopic fractionation could be calculated for their test temperature of 10K at various densities, and for a given initial \ce{^{12}C}:\ce{^{13}C} and \ce{^{16}O}:\ce{^{18}O} abundance ratio. It can be seen in Table~\ref{tab:fractionation} that the abundance ratios are within an order of magnitude of those predicted by \citet{Mladenovic2014}, strongly suggesting fractionation, but differ in detail systematically by a factor of 3 in the $^{13}$C species. 
\begin{table}
  \centering
  \caption{Abundance ratios compared to fractionation calculations (1) Source name (2) $^{13}$CO/C$^{18}$O modelled here (3) $^{13}$CO/C$^{18}$O calculated by \citep{Mladenovic2014} at 10 K (4) HCO$^+$/H$^{13}$CO$^+$ modelled here (5) HCO$^+$/H$^{13}$CO$^+$ calculated by \citet{Mladenovic2014} at 10K. Abundance ratios are within around a factor of 3 of fractionation calculations with differences likely due to the higher temperature in the G333 cores}
  \label{tab:fractionation}
  \begin{tabular}{@{}ccccc}
    \hline
    Source & $^{13}$CO/C$^{18}$O & & HCO$^+$/H$^{13}$CO$^+$ & \\
    $T_{\rm cold}$ & Modelled & MR & Modelled  &  MR \\
    (1) & (2) & (3) & (4) & (5) \\
    \hline
    G333.6--0.2 & 3.0 & 10.5 & 5.0 &  17.6 \\
    G333.1--0.4 & 7.0 & 10.5  & 4.0 &  17.6 \\
    G332.8--0.5 & 3.3 & 10.5 & 8.0 &  17.6 \\
    \hline
  \end{tabular}\\
\end{table}

We can suggest three reasons why our abundance ratios differ somewhat from those of  \citet{Mladenovic2014}. Firstly, our core conditions differ in being warmer than their calculation temperature of 10K. \citet{Mladenovic2014} report the zero point energy differences for the network of fractionation reactions to all be between around 7.5K and 35K, with the principal \ce{^{13}CO} fractionation reaction above having the highest $\Delta E$. At the higher gas temperatures here, the conditions may favour the particular enhancement of $^{13}$CO and H$^{13}$CO$^+$ by suppressing other possible fractionation processes (for example, the exchange reaction H$^{13}$CO$^+$ + $^{13}$C$^{18}$O$\rightarrow$ H$^{13}$C$^{18}$O$^+$ +$^{13}$CO is reversible at the temperature of the G333 cores).  Secondly, though this is a minor factor, the initial atomic abundance ratios used should be lower because of the galactocentric radius of G333; the values in \citet{Mladenovic2014} could be crudely scaled by a factor equal to the ratio of the abundance ratios at the two points i.e. by 50/60 for \ce{^{12}C}:\ce{^{13}C}.  Thirdly, and perhaps most importantly, for the reactions to favour the fractionation of \ce{^{13}C} species, the presence of \ce{^{13}C+} is required, which in turn requires the presence of light ionization. We expect that shocks arising from the supersonic turbulent velocities present in these cores will lead to a higher degree of ionization than used by\citet{Mladenovic2014} who adopted the local interstellar radiation field as the input ionizing flux for their calculations.

\subsection{Implications for the understanding of HMSF} \label{sec:implications}
The {\sc mollie} models show that for these three sources, HMSF is consistent with a scaled-up version (by a factor of ten) in turbulent velocity and infall velocity of LMSF, and thus is not consistent with the coalescence models of star formation. In the LMSF case, the turbulent velocities are subsonic, whereas the scaled-up HMSF case is supersonic. Supersonic speeds in molecular clouds are consistent with collapse via fragmentation and hence the turbulent and clumpy models of star formation. In Table~\ref{tab:velocities}, we summarize the typical velocities found in star formation regions of different mass, with the highest mass objects such as those presented in this paper being consistent with scaled-up versions of the lower mass case as would be expected based on virial arguments. For the case of $10^3$ \solarM, the observed linewidths ($\sim\,7$\,\kms) are wider than virial ($\sim\,2$\,\kms), which suggests that the cores are not in equilibrium, which may be due the fact that the core is collapsing, and/or turbulent motions. This is also consistent with the suggestion of accelerated collapse due to external pressure as discussed in previous section. The 3-D modelling reveals that very different line profile shapes are reproduced by quite similar underlying models in which the orientation of an hourglass-shaped outflow is the principal variation between the sources. This has also been found in recent modelling work of low mass class 0/I sources \citep{Rawlings2013,Carolan2008} and in our previous studies of massive star forming regions \citep{Lo2011, Carolan2009} and further supports the view that the infall, rotation and outflow processes at work in massive star formation are scaled-up versions of those in low mass star formation.

\begin{table}
  \caption{Comparision of velocities found in star forming regions of differing mass scale. Columns are (1) typical mass; (2) measured line width; (3) infall velocity; (4) turbulent velocity; (5) references.}
  \label{tab:velocities}
  \begin{tabular}{@{}ccccc}
    \hline
    $M$ & $\Delta v$ & $v_{\rm infall}$ & $v_{\rm turb}$ & Reference \\
    \solarM & \kms & \kms & \kms & \\
    (1) & (2) & (3) & (4) & (5) \\
    \hline
    $\sim 1$ & 3 & 0.1 & 0.15 & \citet{Carolan2008} \\
    $\sim 800$ & 5 & 1 & 1.5 & \citet{Carolan2009} \\
    $\sim 10^3$ & 7 & 3 & 3.5 & This paper \\
    \hline
  \end{tabular}
\end{table}

\section{SUMMARY} \label{sec:summary}
Data from three massive star-forming cores in the G333 giant molecular cloud, obtained with Mopra was used to investigate the dynamics and chemistry of these sources. Self-consistent 3-D molecular line radiative transfer modelling of the observed lines with the {\sc mollie} code has confirmed the presence of infall in the sources. The modelled infall velocities are 1.5 to 3.5 \kms\ with infall rate of order $1-2 \times 10^{-3}$ \solarM\,yr$^{-1}$. Derived time-scales suggest the three cores may be undergoing accelerated collapse as shown in \citet{Lintott2005}, possibly due to external pressure from dynamical activities among high-mass star forming regions. There is evidence of significant molecular depletion due to freeze-out, with the remaining gas phase molecules being subject to strong fractionation effects in \ce{^{13}C} species. The turbulent and infall velocities associated with these HMSFRs are an order of magnitude larger than those expected in the low-mass case. The highly supersonic velocity structure in turbulence and infall will very naturally lead to localised shocks that could supply the $^{13}$C$^+$ ions needed to fractionate CO and HCO$^+$. The many similarities of these massive cores with their low mass counterparts is supportive of view of massive star formation being a scaled-up version of low mass star formation. The principal and crucial difference is that the internal velocity fields are highly supersonic in the massive star formation case, and this gives a natural way to fragment the cloud into many individual star forming units, as required to form the dense clusters of low mass stars seen associated with young massive stars.

\section*{ACKNOWLEDGEMENTS}
NL's postdoctoral fellowship is supported by a CONICYT/FONDECYT postdoctorado, under project no. 3130540. NL acknowledges partial support from the ALMA-CONICYT Fund for the Development of Chilean Astronomy Project 31090013, Center of Excellence in Astrophysics and Associated Technologies (PFB 06) and Centro de Astrof\'{i}sica FONDAP\,15010003. MPR acknowledges support from a Science Foundation Ireland Research Frontiers grant. LB acknowledges support from CONICYT Project PFB 06. The Mopra Telescope and ATCA are part of the Australia Telescope and are funded by the Commonwealth of Australia for operation as National Facility managed by CSIRO. The UNSW-MOPS Digital Filter Bank used for the observations with the Mopra Telescope was provided with support from the University of New South Wales, Monash University, University of Sydney and Australian Research Council. This work has made use of the SIMBAD database, operated at CDS, Strasbourg, France and also NASA's Astrophysics Data System. We have also made use of the NIST Recommended Rest Frequencies for Observed Interstellar Molecular Microwave Transitions, by Frank J. Lovas. We thank the referee for the constructive comments on improving this work.

\bibliographystyle{mn2e}
%\bibliography{infalloutflow}

\begin{thebibliography}{}

\bibitem[\protect\citeauthoryear{Alves, Lada \& Lada}{Alves
  et~al.}{2001}]{Alves01}
Alves J.~F.,  Lada C.~J.,    Lada E.~A.,  2001, Nature, 409, 159

\bibitem[\protect\citeauthoryear{{Arce} \& {Goodman}}{{Arce} \&
  {Goodman}}{2002}]{arce_goodman02}
{Arce} H.~G.,  {Goodman} A.~A.,  2002, ApJ, 575, 911

\bibitem[\protect\citeauthoryear{{Arreaga-Garc{\'{\i}}a}, {Klapp-Escribano} \&
  {G{\'o}mez-Ram{\'{\i}}rez}}{{Arreaga-Garc{\'{\i}}a}
  et~al.}{2010}]{arreaga-garcia.et.al10}
{Arreaga-Garc{\'{\i}}a} G.,  {Klapp-Escribano} J.,
  {G{\'o}mez-Ram{\'{\i}}rez} F.,  2010, A\&A, 509, A96

\bibitem[\protect\citeauthoryear{{Bains}, {Wong}, {Cunningham}, {Sparks},
  {Brisbin}, {Calisse}, {Dempsey}, {Deragopian} et~al.,}{{Bains}
  et~al.}{2006}]{bains06}
{Bains} I.,  {Wong} T.,  {Cunningham} M.,  {Sparks} P.,  {Brisbin} D.,
  {Calisse} P.,  {Dempsey} J.~T.,  {Deragopian} G.,    et~al., 2006, MNRAS,
  367, 1609

\bibitem[\protect\citeauthoryear{Bonner}{Bonner}{1956}]{bonner56}
Bonner W.~B.,  1956, MNRAS, 116, 351

\bibitem[\protect\citeauthoryear{{Brinch} \& {Hogerheijde}}{{Brinch} \&
  {Hogerheijde}}{2010}]{brinch_hogerheijde2010}
{Brinch} C.,  {Hogerheijde} M.~R.,  2010, A\&A, 523, A25

\bibitem[\protect\citeauthoryear{{Burton}, {Braiding}, {Glueck}, {Goldsmith},
  {Hawkes}, {Hollenbach}, {Kulesa}, {Martin}, {Pineda}, {Rowell}, {Simon},
  {Stark}, {Stutzki}, {Tothill}, {Urquhart}, {Walker}, {Walsh} \&
  {Wolfire}}{{Burton} et~al.}{2013}]{burton13}
{Burton} M.~G.,  {Braiding} C.,  {Glueck} C.,  {Goldsmith} P.,  {Hawkes} J.,
  {Hollenbach} D.~J.,  {Kulesa} C.,  {Martin} C.~L.,  {Pineda} J.~L.,  {Rowell}
  G.,  {Simon} R.,  {Stark} A.~A.,  {Stutzki} J.,  {Tothill} N.~J.~H.,
  {Urquhart} J.~S.,  {Walker} C.,  {Walsh} A.~J.,    {Wolfire} M.,  2013,
  PASA, 30, 44

\bibitem[\protect\citeauthoryear{{Carolan}, {Khanzadyan}, {Redman}, {Thompson},
  {Jones}, {Cunningham}, {Loughnane}, {Bains} et~al.,}{{Carolan}
  et~al.}{2009}]{Carolan2009}
{Carolan} P.~B.,  {Khanzadyan} T.,  {Redman} M.~P.,  {Thompson} M.~A.,  {Jones}
  P.~A.,  {Cunningham} M.~R.,  {Loughnane} R.~M.,  {Bains} I.,    et~al., 2009,
  MNRAS, 400, 78

\bibitem[\protect\citeauthoryear{{Carolan}, {Redman}, {Keto} \&
  {Rawlings}}{{Carolan} et~al.}{2008}]{Carolan2008}
{Carolan} P.~B.,  {Redman} M.~P.,  {Keto} E.,    {Rawlings} J.~M.~C.,  2008,
  MNRAS, 383, 705

\bibitem[\protect\citeauthoryear{{Chandrasekhar}}{{Chandrasekhar}}{1967}]{chandrasekhar67}
{Chandrasekhar} S.,  1967, {An introduction to the study of stellar structure}.
New York: Dover, 1967

\bibitem[\protect\citeauthoryear{Christie et al.}{2012}]{christie12} Christie H., Viti S., Yates J., Hatchell J., Fuller G. A., Duarte-Cabral A., Sadavoy S., Buckle J. V., et al., 2012, MNRAS, 422, 968 

\bibitem[\protect\citeauthoryear{Choi, {Evans II}, Gregersen \& Wang}{Choi
  et~al.}{1995}]{Choi95}
Choi M.,  {Evans II} N.~J.,  Gregersen E.~M.,    Wang Y.,  1995, ApJ, 448, 742

\bibitem[\protect\citeauthoryear{{Ebert}}{{Ebert}}{1955}]{ebert55}
{Ebert} R.,  1955, Zeitschrift fur Astrophysics, 37, 217

\bibitem[\protect\citeauthoryear{{Garc{\'{\i}}a}, {Bronfman}, {Nyman}, {Dame}
  \& {Luna}}{{Garc{\'{\i}}a} et~al.}{2014}]{garcia14}
{Garc{\'{\i}}a} P.,  {Bronfman} L.,  {Nyman} L.-{\AA}.,  {Dame} T.~M.,
  {Luna} A.,  2014, ApJS, 212, 2

\bibitem[\protect\citeauthoryear{{Hirota}, {Yamamoto}, {Mikami} \&
  {Ohishi}}{{Hirota} et~al.}{1998}]{Hirota1998}
{Hirota} T.,  {Yamamoto} S.,  {Mikami} H.,    {Ohishi} M.,  1998, ApJ, 503, 717

\bibitem[\protect\citeauthoryear{Hodapp}{Hodapp}{1998}]{hodapp98}
Hodapp K.~W.,  1998, ApJ, 500, L183

\bibitem[\protect\citeauthoryear{{Hogerheijde} \& {van der Tak}}{{Hogerheijde}
  \& {van der Tak}}{2000}]{hogerheijde_vandertak2000}
{Hogerheijde} M.~R.,  {van der Tak} F.~F.~S.,  2000, A\&A, 362, 697

\bibitem[\protect\citeauthoryear{Jackson et al.}{2013}]{jackson13} 
Jackson J. M., Rathborne J. M., Foster J. B., Whitaker J. S., Sanhueza P., Claysmith C., Mascoop J. L., Wienen M., et al., 2013, PASA, 30, 
57 

\bibitem[\protect\citeauthoryear{{Jones}, {Cunningham}, {Bains}, {Muller},
  {Wong} \& {Burton}}{{Jones} et~al.}{2007}]{bains07}
{Jones} P.~A.,  {Cunningham} M.~R.,  {Bains} I.,  {Muller} E.,  {Wong} T.,
  {Burton} M.~G.,  2007, in {Elmegreen} B.~G.,  {Palous} J.,  eds, IAU
  Symposium Vol.~237 of IAU Symposium, {Turbulence in the G333 molecular
  cloud}.
pp 429--429

\bibitem[\protect\citeauthoryear{{Jordan}, {Walsh}, {Lowe}, {Voronkov},
  {Ellingsen}, {Breen}, {Purcell}, {Barnes}, {Burton}, {Cunningham}, {Hill},
  {Jackson}, {Longmore}, {Peretto} \& {Urquhart}}{{Jordan}
  et~al.}{2015}]{jordan15}
{Jordan} C.~H.,  {Walsh} A.~J.,  {Lowe} V.,  {Voronkov} M.~A.,  {Ellingsen}
  S.~P.,  {Breen} S.~L.,  {Purcell} C.~R.,  {Barnes} P.~J.,  {Burton} M.~G.,
  {Cunningham} M.~R.,  {Hill} T.,  {Jackson} J.~M.,  {Longmore} S.~N.,
  {Peretto} N.,    {Urquhart} J.~S.,  2015, MNRAS, 448, 2344

\bibitem[\protect\citeauthoryear{{J{\o}rgensen}, {Bourke}, {Myers}, {Di
  Francesco}, {van Dishoeck}, {Lee}, {Ohashi}, {Sch{\"o}ier}, {Takakuwa},
  {Wilner} \& {Zhang}}{{J{\o}rgensen} et~al.}{2007}]{jorgensen.et.al07}
{J{\o}rgensen} J.~K.,  {Bourke} T.~L.,  {Myers} P.~C.,  {Di Francesco} J.,
  {van Dishoeck} E.~F.,  {Lee} C.-F.,  {Ohashi} N.,  {Sch{\"o}ier} F.~L.,
  {Takakuwa} S.,  {Wilner} D.~J.,    {Zhang} Q.,  2007, ApJ, 659, 479

\bibitem[\protect\citeauthoryear{{Keto}, {Rybicki}, {Bergin} \& {Plume}}{{Keto}
  et~al.}{2004}]{Keto2004}
{Keto} E.,  {Rybicki} G.~B.,  {Bergin} E.~A.,    {Plume} R.,  2004, ApJ, 613,
  355

\bibitem[\protect\citeauthoryear{{Langer} \& {Penzias}}{{Langer} \&
  {Penzias}}{1990}]{Langer1990}
{Langer} W.~D.,  {Penzias} A.~A.,  1990, ApJ, 357, 477

\bibitem[\protect\citeauthoryear{{Lee}, {Mundy}, {Stone} \& {Ostriker}}{{Lee}
  et~al.}{2002}]{lee.et.al02}
{Lee} C.-F.,  {Mundy} L.~G.,  {Stone} J.~M.,    {Ostriker} E.~C.,  2002, ApJ,
  576, 294

\bibitem[\protect\citeauthoryear{{Lintott}, {Viti}, {Rawlings}, {Williams},
  {Hartquist}, {Caselli}, {Zinchenko} \& {Myers}}{{Lintott}
  et~al.}{2005}]{Lintott2005}
{Lintott} C.~J.,  {Viti} S.,  {Rawlings} J.~M.~C.,  {Williams} D.~A.,
  {Hartquist} T.~W.,  {Caselli} P.,  {Zinchenko} I.,    {Myers} P.,  2005, ApJ,
  620, 795

%\bibitem[\protect\citeauthoryear{{Lo}, {Cunningham}, {Bains}, {Burton} \&
%  {Garay}}{{Lo} et~al.}{2007a}]{lo07}
%{Lo} N.,  {Cunningham} M.,  {Bains} I.,  {Burton} M.~G.,    {Garay} G.,  2007a,
%  MNRAS, 381, L30

\bibitem[\protect\citeauthoryear{{Lo}, {Cunningham}, {Bains}, {Burton} \&
  {Garay}}{{Lo} et~al.}{2007b}]{Lo2007}
{Lo} N.,  {Cunningham} M.,  {Bains} I.,  {Burton} M.~G.,    {Garay} G.,  2007b,
  MNRAS, 381, L30

\bibitem[\protect\citeauthoryear{{Lo}, {Cunningham}, {Jones}, {Bains},
  {Burton}, {Wong}, {Muller}, {Kramer} et~al.,}{{Lo} et~al.}{2009}]{Lo2009}
{Lo} N.,  {Cunningham} M.~R.,  {Jones} P.~A.,  {Bains} I.,  {Burton} M.~G.,
  {Wong} T.,  {Muller} E.,  {Kramer} C.,    et~al., 2009, MNRAS, 395, 1021

%\bibitem[\protect\citeauthoryear{{Lo}, {Cunningham}, {Jones}, {Bains},
%  {Burton}, {Wong}, {Muller}, {Kramer}, {Ossenkopf}, {Henkel}, {Deragopian},
%  {Donnelly} \& {Ladd}}{{Lo} et~al.}{2009}]{lo09}
%{Lo} N.,  {Cunningham} M.~R.,  {Jones} P.~A.,  {Bains} I.,  {Burton} M.~G.,
%  {Wong} T.,  {Muller} E.,  {Kramer} C.,  {Ossenkopf} V.,  {Henkel} C.,
%  {Deragopian} G.,  {Donnelly} S.,    {Ladd} E.~F.,  2009, MNRAS, 395, 1021

%\bibitem[\protect\citeauthoryear{{Lo}, {Redman}, {Jones}, {Cunningham},
%  {Chhetri}, {Bains} \& {Burton}}{{Lo} et~al.}{2011a}]{lo11}
%{Lo} N.,  {Redman} M.~P.,  {Jones} P.~A.,  {Cunningham} M.~R.,  {Chhetri} R.,
%  {Bains} I.,    {Burton} M.~G.,  2011a, MNRAS, 415, 525

\bibitem[\protect\citeauthoryear{{Lo}, {Redman}, {Jones}, {Cunningham},
  {Chhetri}, {Bains} \& {Burton}}{{Lo} et~al.}{2011}]{Lo2011}
{Lo} N.,  {Redman} M.~P.,  {Jones} P.~A.,  {Cunningham} M.~R.,  {Chhetri} R.,
  {Bains} I.,    {Burton} M.~G.,  2011, MNRAS, 415, 525

\bibitem[\protect\citeauthoryear{{Lo}, {Wiles}, {Redman}, {Cunningham},
  {Bains}, {Jones}, {Burton} \& {Bronfman}}{{Lo} et~al.}{2015}]{LoObs}
{Lo} N.,  {Wiles} B.,  {Redman} M.~P.,  {Cunningham} M.~R.,  {Bains} I.,
  {Jones} P.~A.,  {Burton} M.~G.,    {Bronfman} L.,  2015, MNRAS, 453, 3245

\bibitem[\protect\citeauthoryear{{Loughnane}, {Redman}, {Thompson}, {Lo},
  {O'Dwyer} \& {Cunningham}}{{Loughnane} et~al.}{2012}]{Loughnane2012}
{Loughnane} R.~M.,  {Redman} M.~P.,  {Thompson} M.~A.,  {Lo} N.,  {O'Dwyer} B.,
     {Cunningham} M.~R.,  2012, MNRAS, 420, 1367

\bibitem[\protect\citeauthoryear{{Milam}, {Savage}, {Brewster}, {Ziurys} \&
  {Wyckoff}}{{Milam} et~al.}{2005}]{Milam2005}
{Milam} S.~N.,  {Savage} C.,  {Brewster} M.~A.,  {Ziurys} L.~M.,    {Wyckoff}
  S.,  2005, ApJ, 634, 1126

\bibitem[\protect\citeauthoryear{{Mladenovi{\'c}} \& {Roueff}}{{Mladenovi{\'c}}
  \& {Roueff}}{2014}]{Mladenovic2014}
{Mladenovi{\'c}} M.,  {Roueff} E.,  2014, A\&A, 566, A144

\bibitem[\protect\citeauthoryear{{Mookerjea}, {Kramer}, {Nielbock} \&
  {Nyman}}{{Mookerjea} et~al.}{2004}]{Mookerjea2004}
{Mookerjea} B.,  {Kramer} C.,  {Nielbock} M.,    {Nyman} L.-{\AA}.,  2004,
  A\&A, 426, 119

\bibitem[\protect\citeauthoryear{{Myers}, {Mardones}, {Tafalla}, {Williams} \&
  {Wilner}}{{Myers} et~al.}{1996}]{Myers1996}
{Myers} P.~C.,  {Mardones} D.,  {Tafalla} M.,  {Williams} J.~P.,    {Wilner}
  D.~J.,  1996, ApJ Lett., 465, L133+

\bibitem[\protect\citeauthoryear{{Qi}, {{\"O}berg} \& {Wilner}}{{Qi}
  et~al.}{2013}]{qi13}
{Qi} C.,  {{\"O}berg} K.~I.,    {Wilner} D.~J.,  2013, ApJ, 765, 34

\bibitem[\protect\citeauthoryear{{Rawlings}, {Hartquist}, {Menten} \&
  {Williams}}{{Rawlings} et~al.}{1992}]{Rawlings1992}
{Rawlings} J.~M.~C.,  {Hartquist} T.~W.,  {Menten} K.~M.,    {Williams} D.~A.,
  1992, MNRAS, 255, 471

\bibitem[\protect\citeauthoryear{{Rawlings}, {Redman} \& {Carolan}}{{Rawlings}
  et~al.}{2013}]{Rawlings2013}
{Rawlings} J.~M.~C.,  {Redman} M.~P.,    {Carolan} P.~B.,  2013, MNRAS, 435,
  289

\bibitem[\protect\citeauthoryear{{Rawlings}, {Redman}, {Keto} \&
  {Williams}}{{Rawlings} et~al.}{2004}]{Rawlings2004}
{Rawlings} J.~M.~C.,  {Redman} M.~P.,  {Keto} E.,    {Williams} D.~A.,  2004,
  MNRAS, 351, 1054

\bibitem[\protect\citeauthoryear{{Redman}, {Keto} \& {Rawlings}}{{Redman}
  et~al.}{2006}]{Redman2006}
{Redman} M.~P.,  {Keto} E.,    {Rawlings} J.~M.~C.,  2006, MNRAS, 370, L1

\bibitem[\protect\citeauthoryear{Redman, Keto, Rawlings \& Williams}{Redman
  et~al.}{2004}]{Redman04}
Redman M.~P.,  Keto E.,  Rawlings J. M.~C.,    Williams D.~A.,  2004, MNRAS,
  352, 1365

\bibitem[\protect\citeauthoryear{{Redman}, {Viti}, {Cau} \&
  {Williams}}{{Redman} et~al.}{2003}]{Redman2003}
{Redman} M.~P.,  {Viti} S.,  {Cau} P.,    {Williams} D.~A.,  2003, MNRAS, 345,
  1291

\bibitem[\protect\citeauthoryear{{Rollins}, {Rawlings}, {Williams} \&
  {Redman}}{{Rollins} et~al.}{2014}]{Rollins14}
{Rollins} R.~P.,  {Rawlings} J.~M.~C.,  {Williams} D.~A.,    {Redman} M.~P.,
  2014, MNRAS, 443, 3033

\bibitem[\protect\citeauthoryear{{Rybicki} \& {Hummer}}{{Rybicki} \&
  {Hummer}}{1991}]{Rybicki1991}
{Rybicki} G.~B.,  {Hummer} D.~G.,  1991, A\&A, 245, 171

\bibitem[\protect\citeauthoryear{{Tafalla}, {Myers}, {Caselli} \&
  {Walmsley}}{{Tafalla} et~al.}{2004}]{Tafalla2004}
{Tafalla} M.,  {Myers} P.~C.,  {Caselli} P.,    {Walmsley} C.~M.,  2004, A\&A,
  416, 191

\bibitem[\protect\citeauthoryear{{van Zadelhoff}, {Dullemond}, {van der Tak},
  {Yates}, {Doty}, {Ossenkopf}, {Hogerheijde}, {Juvela} et~al.,}{{van
  Zadelhoff} et~al.}{2002}]{van_Zadelhoff2002}
{van Zadelhoff} G.-J.,  {Dullemond} C.~P.,  {van der Tak} F.~F.~S.,  {Yates}
  J.~A.,  {Doty} S.~D.,  {Ossenkopf} V.,  {Hogerheijde} M.~R.,  {Juvela} M.,
  et~al., 2002, A\&A, 395, 373

\bibitem[\protect\citeauthoryear{Walsh et al.}{2011}]{walsh11} 
Walsh A.~J., et al., 2011, MNRAS, 416, 1764 

\bibitem[\protect\citeauthoryear{Ward-Thompson \& Buckley}{Ward-Thompson \&
  Buckley}{2001}]{ward_thompson_buckley01}
Ward-Thompson D.,  Buckley H.~D.,  2001, MNRAS, 327, 955

\bibitem[\protect\citeauthoryear{{Whitworth} \& {Ward-Thompson}}{{Whitworth} \&
  {Ward-Thompson}}{2001}]{whitworth01}
{Whitworth} A.~P.,  {Ward-Thompson} D.,  2001, ApJ, 547, 317

\bibitem[\protect\citeauthoryear{{Wong}, {Ladd}, {Brisbin}, {Burton}, {Bains},
  {Cunningham}, {Lo}, {Jones} et~al.,}{{Wong} et~al.}{2008}]{wong08}
{Wong} T.,  {Ladd} E.~F.,  {Brisbin} D.,  {Burton} M.~G.,  {Bains} I.,
  {Cunningham} M.~R.,  {Lo} N.,  {Jones} P.~A.,    et~al., 2008, MNRAS, 386,
  1069

\bibitem[\protect\citeauthoryear{{Wu}, {Zhu}, {Wei}, {Xu}, {Zhang} \&
  {Fiege}}{{Wu} et~al.}{2005}]{Wu2005}
{Wu} Y.,  {Zhu} M.,  {Wei} Y.,  {Xu} D.,  {Zhang} Q.,    {Fiege} J.~D.,  2005,
  ApJ Lett., 628, L57

\bibitem[\protect\citeauthoryear{Zhou, {Evans II}, K{\"{o}mpe} \&
  Walmsley}{Zhou et~al.}{1993}]{zhou.et.al93}
Zhou S.,  {Evans II} N.~J.,  K{\"{o}mpe} C., Walmsley C.~M.,  1993, ApJ,
  404, 232


\end{thebibliography}

%\appendix
%\section{appendix section}

\bsp

\label{lastpage}

\end{document}